\documentclass[iop, numberedappendix]{emulateapj}

\usepackage{apjfonts}

\usepackage{color}
\definecolor{citeRGB}{rgb}{0,0.1,0.7}
\usepackage[hyperfootnotes=true,naturalnames=true,letterpaper,pdfstartview=FitH,pdfpagemode=UseNone,colorlinks=true,citecolor=citeRGB]{hyperref}

\usepackage{etoolbox}

\makeatletter

\patchcmd{\NAT@citex}
  {\@citea\NAT@hyper@{\NAT@nmfmt{\NAT@nm}\NAT@date}}
  {\@citea\NAT@nmfmt{\NAT@nm}\NAT@hyper@{\NAT@date}}
  {}
  {}

\patchcmd{\NAT@citex}
  {\@citea\NAT@hyper@{%
     \NAT@nmfmt{\NAT@nm}%
     \hyper@natlinkbreak{\NAT@aysep\NAT@spacechar}{\@citeb\@extra@b@citeb}%
     \NAT@date}}
  {\@citea\NAT@nmfmt{\NAT@nm}%
   \NAT@aysep\NAT@spacechar%
   \NAT@hyper@{\NAT@date}}
  {}
  {}

\patchcmd{\NAT@citex}
  {\@citea\NAT@hyper@{%
     \NAT@nmfmt{\NAT@nm}%
     \hyper@natlinkbreak{\NAT@spacechar\NAT@@open\if*#1*\else#1\NAT@spacechar\fi}%
       {\@citeb\@extra@b@citeb}%
     \NAT@date}}
  {\@citea\NAT@nmfmt{\NAT@nm}%
   \NAT@spacechar\NAT@@open\if*#1*\else#1\NAT@spacechar\fi%
   \NAT@hyper@{\NAT@date}}
  {}
  {}

\makeatother

\gdef\HST{\textit{HST}}
\gdef\G141{\textit{G141}}
\gdef\F140W{\textit{F140W}}
\gdef\fluxcgs{\mathrm{erg\ s^{-1}\ cm^{-2}}}
\gdef\micront{$\mu$m}

\gdef\aXe{\textsc{aXe}}
\gdef\flux_radius{\textsc{flux\_radius}}
\gdef\epers{\textit{e}$^{-}$ s$^{-1}$}

\gdef\peryr{\mathrm{yr}^{-1}}
\gdef\kms{km\,s$^{-1}$}

\gdef\24mum{$24\,\mu\mathrm{m}$}
\gdef\arcsec{^{\prime\prime}}

\gdef\4ang{4000\,\AA}

\gdef\eazy{\textsc{Eazy}}

\gdef\micron{\mbox{$\mu\mathrm{m}$}}

\shortauthors{Brammer et al.}
\shorttitle{3D-HST grism survey}
\slugcomment{ApJS Submitted, \today}

\begin{document}

\title{\textit{3D-HST}: A wide-field grism spectroscopic survey with the \textit{Hubble Space Telescope}\footnotemark[*]}

\author{Gabriel~B.~Brammer\altaffilmark{1}, 
Pieter G.\ van Dokkum\altaffilmark{2},
Marijn Franx\altaffilmark{3},
Mattia Fumagalli\altaffilmark{3},
Shannon Patel\altaffilmark{3},
Hans-Walter Rix\altaffilmark{4},
Rosalind E.\ Skelton\altaffilmark{2},
Mariska Kriek\altaffilmark{5},
Erica Nelson\altaffilmark{2},
Kasper B.\ Schmidt\altaffilmark{4},
Rachel Bezanson\altaffilmark{2},
Elisabete da Cunha\altaffilmark{4},
Dawn K.\ Erb\altaffilmark{7},
Xiaohui Fan\altaffilmark{6},
Natascha F\"orster Schreiber\altaffilmark{8},
Garth D.\ Illingworth\altaffilmark{9},
Ivo Labb\'e\altaffilmark{3},
Joel Leja\altaffilmark{2},
Britt Lundgren\altaffilmark{2},
Dan Magee\altaffilmark{9},
Danilo Marchesini\altaffilmark{10},
Patrick McCarthy\altaffilmark{11},
Ivelina Momcheva\altaffilmark{11,2},
Adam Muzzin\altaffilmark{3},
Ryan Quadri\altaffilmark{11},
Charles C.\ Steidel\altaffilmark{12},
Tomer Tal\altaffilmark{2},
David Wake\altaffilmark{2},
Katherine E.\ Whitaker\altaffilmark{2},
Anna Williams\altaffilmark{13}
}

\email{gbrammer@eso.org}

\altaffiltext{1}
{European Southern Observatory, Alonso de C\'ordova 3107, Casilla 19001, Vitacura, Santiago, Chile}
\altaffiltext{2}
{Department of Astronomy, Yale University, New Haven, CT 06520, USA}
\altaffiltext{3}
{Leiden Observatory, Leiden University, Leiden, The Netherlands}
\altaffiltext{4}
{Max Planck Institute for Astronomy (MPIA), K\"onigstuhl 17,
69117, Heidelberg, Germany}
\altaffiltext{5}{Harvard-Smithsonian Center for Astrophysics,
60 Garden Street, Cambridge, MA 02138, USA}
\altaffiltext{6}{Steward Observatory,
University of Arizona, Tucson, AZ 85721, USA}
\altaffiltext{7}
{Department of Physics, University of Wisconsin-Milwaukee, P.O. Box 413,
Milwaukee, WI 53201, USA}
\altaffiltext{8}
{Max-Planck-Institut f\"ur extraterrestrische Physik,
Giessenbachstrasse, D-85748 Garching, Germany}
\altaffiltext{9}
{Astronomy Department, University of California, Santa Cruz, CA 95064, USA}
\altaffiltext{10}
{Physics and Astronomy Department, Tufts University, Robinson Hall,
Room 257, Medford, MA, 02155, USA}
\altaffiltext{11}
{Carnegie Observatories, 813 Santa Barbara Street, Pasadena, CA 91101, USA}
\altaffiltext{12}
{Department of Astronomy, California Institute of Technology, MS 249-17, Pasadena, CA 91125, USA}
\altaffiltext{13}
{Department of Astronomy, University of Wisconsin-Madison, 475
North Charter Street, Madison, WI 53706, USA}

\footnotetext[*]{Based on observations made with the NASA/ESA \textit{Hubble Space Telescope}, obtained at the Space Telescope Science Institute, which is operated by the Association of Universities for Research in Astronomy, Inc., under NASA contract NAS 5-26555. These observations are associated with programs \#12177, 12328.}


\begin{abstract}

We present 3D-HST, a near-infrared spectroscopic Treasury program with the \textit{Hubble Space Telescope} for studying the physical processes that shape galaxies in the distant universe.  3D-HST provides rest-frame optical spectra for a sample of $\sim$7000 galaxies at $1<z<3.5$, the epoch when $\sim$60\% of all star formation took place, the number density of quasars peaked, the first galaxies stopped forming stars, and the structural regularity that we see in galaxies today must have emerged.  3D-HST will cover three-quarters (625 arcmin$^2$) of the CANDELS Treasury survey area with two orbits of primary WFC3/G141 grism coverage and two to four orbits with the ACS/G800L grism in parallel.  In the IR these exposure times yield a continuum signal-to-noise of $\sim$5 per resolution element at $H_{140}\sim23.1$ and a 5$\sigma$ emission line sensitivity of $\sim5\times10^{-17}\ \fluxcgs$ for typical objects, improving by a factor of $\sim2$ for compact sources in images with low sky background levels.  	The WFC3/G141 spectra provide continuous wavelength coverage from 1.1 to 1.6 \micront\ at a spatial resolution of $\sim0\farcs13$, which, combined with their depth, makes them a unique resource for studying galaxy evolution.  We present an overview of the preliminary reduction and analysis of the grism observations, including emission line and redshift measurements from combined fits to the extracted grism spectra and photometry from ancillary multi-wavelength catalogs.  The present analysis yields redshift estimates with a precision of $\sigma(z)=0.0034(1+z)$, or $\sigma(v)\approx 1000$ \kms.  We illustrate how the generalized nature of the survey yields near-infrared spectra of remarkable quality for many different types of objects, including a quasar at $z=4.7$, quiescent galaxies at $z\sim2$, and the most distant T-type brown dwarf star known.  The combination of the CANDELS and 3D-HST surveys will provide the definitive imaging and spectroscopic dataset for studies of the $1 < z < 3.5$ universe until the launch of the \textit{James Webb Space Telescope}.

\end{abstract}


%
%
\section{Introduction}\label{s:intro}

The investment of thousands of orbits of \textit{Hubble Space Telescope} (\HST) time has provided an incomparable imaging legacy that has revolutionized our understanding of observational cosmology and galaxy formation.  The \HST's location in low-Earth orbit enables high-spatial resolution free from the distorting effects of the atmosphere and remarkable sensitivity, particularly at near-infrared (IR) wavelengths, due to much lower background levels compared to the ground.  As but two examples, these unique capabilities have helped confirm the discovery of the accelerating expansion of the universe \citep{riess:04} and helped to establish that the epoch at $1 < z < 3$ is a critical period for the evolution of galaxies, when $\sim$60\% of the cosmic star formation took place \citep[e.g.,][]{ahopkins:06, bouwens:07} and the structural regularity of galaxies seen today emerged \citep[e.g.,][]{elmegreen:07, wuyts:11b}

The interpretation of deep, high spatial resolution \HST\ images of galaxies at $z>1$ is typically limited by the lack of the crucial third dimension, redshift, and other physical diagnostics that can be measured from the galaxies' spectra such as their star formation rates (SFRs) and metallicities.  Large spectroscopic surveys selected at optical wavelengths are typically limited to $z\lesssim1$ for magnitude-limited samples \citep[e.g., zCOSMOS,][]{lilly:07} or to color-selected, UV-luminous galaxies at $z>2$ \citep[e.g.,][]{steidel:99, steidel:03}, which represent only a biased subset of the $M>M_*$ galaxy population at these redshifts \citep{vandokkum:06}.  Representative spectroscopic samples of galaxies at $z > 1$ must be observed in the near-IR, which contains redshifted light emitted primarily by longer-lived stars \citep[e.g.,][]{maraston:10, wuyts:12} as well as well-calibrated spectral features such as the H$\alpha$ and [\ion{O}{3}] emission lines \citep[e.g.,][]{erb:06}.  However, the high sky background at these wavelengths makes near-IR spectroscopic surveys expensive and has limited sample sizes to the order of a few dozen to $\sim$100 galaxies at $z>1.5$ \citep[e.g.,][]{kriek:08a, forster:09, law:09, gnerucci:11, mancini:11}.  In order to measure the statistical properties of the full galaxy population at $z>1$, IR-selected photometric surveys rely on redshifts and crude spectral diagnostics estimated from multi-wavelength sampling of galaxy spectral energy distributions (SEDs) with broad-band or medium-band filters \citep[e.g.,][]{ilbert:10, brammer:11}.  Current state-of-the-art photometric techniques provide a redshift precision of $\sigma/(1+z)\sim 2\%$ \citep[e.g.,][]{whitaker:11} and stellar mass estimates with precision $\sim$0.1 dex \citep{taylor:11}, but with potentially large systematic uncertainties \citep[e.g.,][]{marchesini:09}.  These photometric measurements are insufficiently precise to study the detailed properties of individual galaxies and their local environment:  the photometric analyses can suffer from type-dependent systematics and even the best photometric redshift measurements still correspond to an uncertainty of $\sim$60~$h^{-1}$ Mpc at $z=2$, many times the correlation length of $M_*$ galaxies at this redshift \citep[cf. $\sim$11~Mpc;][]{wake:11}.

Grism spectroscopy from space provides a promising bridge between ground-based photometric and spectroscopic surveys, combining the depth and multiplexing capabilities of the former with the spectral diagnostics of the latter.  Using the \HST-NICMOS G141 grism spanning the $J$ and $H$ bands, \cite{mccarthy:99} detected H$\alpha$ emission lines at $0.75<z<1.9$, free from the limitations imposed by the near-IR atmospheric absorption and OH emission features.  \cite{yan:99} and \cite{shim:09} use H$\alpha$ observed with the NICMOS grism to estimate the SFR density of the universe out to $z=1.9$.  The line sensitivity of the early NICMOS grism observations is similar to that of wide-field narrow-band photometric surveys \citep[e.g., HIZELS $z=2.3$, ][]{geach:08}, but the grism observations sample similar or larger cosmic volumes for even a limited number of pointings due to the fact that they cover a much broader range of redshifts.  At optical wavelengths, deep observations with the Advanced Camera for Surveys (ACS) G800L grism have been used to confirm the redshifts of $4 < z < 7$ Lyman-break and Ly$\alpha$-emitting galaxies \citep{pirzkal:04, rhoads:09}, to confirm the redshifts of passive galaxies at $\langle z\rangle=1.7$ from rest-frame UV spectral features \citep{daddi:05}, and to study low-mass line emitting galaxies at $0 < z < 2$ \citep{xu:07, straughn:09} at magnitudes fainter than those reached by typical ground-based spectroscopic surveys.  For large-area surveys, however, it is difficult for the ACS grism to compete with ground-based instruments, as much larger telescope apertures are available on the ground where the sky background at optical wavelengths is not as much of a limiting factor.

The improved sensitivity and larger field size ($\sim2\times2$\ arcmin) of the Wide-field Camera 3 (WFC3), installed on the \HST\ in 2009, increases the survey speed by a factor of $\sim$20 compared to the NICMOS grism \citep[][]{atek:10} and enables a true wide-field near-IR spectroscopic survey with the \HST\ that would currently be infeasible from the ground.  The first single science pointing with the WFC3 grism, taken as part of the  Early Release Science program (PI: O'Connell, GO-11359),  demonstrated the remarkable capabilities of this mode:  \cite{vandokkum:10b} used the Balmer absorption features in the spectrum of a single extremely massive, quiescent galaxy at $z=1.9$ to measure a precise redshift and put strong constraints on its star formation history, and \cite{straughn:11} measured redshifts for 48 emission line galaxies at $0.2 < z < 2.3$ down to $m_\mathrm{F098M}=26.9$.  \cite{atek:10} present preliminary results of the WISP parallel survey with the WFC3 G102 and G141 grisms, showing their power for detecting faint emission lines over an extended survey area.  

In this paper we present the 3D-HST survey, an \HST\ Treasury program that is providing WFC3/IR primary and ACS/WFC parallel imaging and grism spectroscopy over $\sim$625 arcmin$^2$ of well-studied extragalactic survey fields.  3D-HST is designed to complement the deep, multi-epoch WFC3 and ACS imaging of the large multi-cycle ``Cosmic Assembly Near-infrared Deep Extragalactic Legacy Survey'' (CANDELS)\footnote{\scriptsize{\url{http://candels.ucolick.org/}}} \citep{grogin:11, koekemoer:11}, by providing spatially-resolved rest-frame optical spectra of $L>0.5L_*$ galaxies over $0.5 < z < 3.5$.  \cite{vandokkum:11} present the first science results from 3D-HST, demonstrating a robust and remarkable diversity within a complete sample of massive ($>10^{11}~M_\odot$) galaxies at $1 < z < 1.5$ thanks to the unique resolved rest-frame optical imaging and H$\alpha$ spectroscopy at these redshifts. In the sections below, we describe the observation strategy, pointing layouts, current data reduction and analysis, and sensitivity of the survey.  We present examples of some noteworthy spectra that illustrate the capabilities of 3D-HST and we conclude with a discussion of the core science goals of the survey.  Magnitudes listed throughout are given in the AB system.

%
%
\section{Observations}\label{s:observations}

\subsection{Filters and grisms}\label{s:filters}

\begin{figure*}
\epsscale{1.}
\plotone{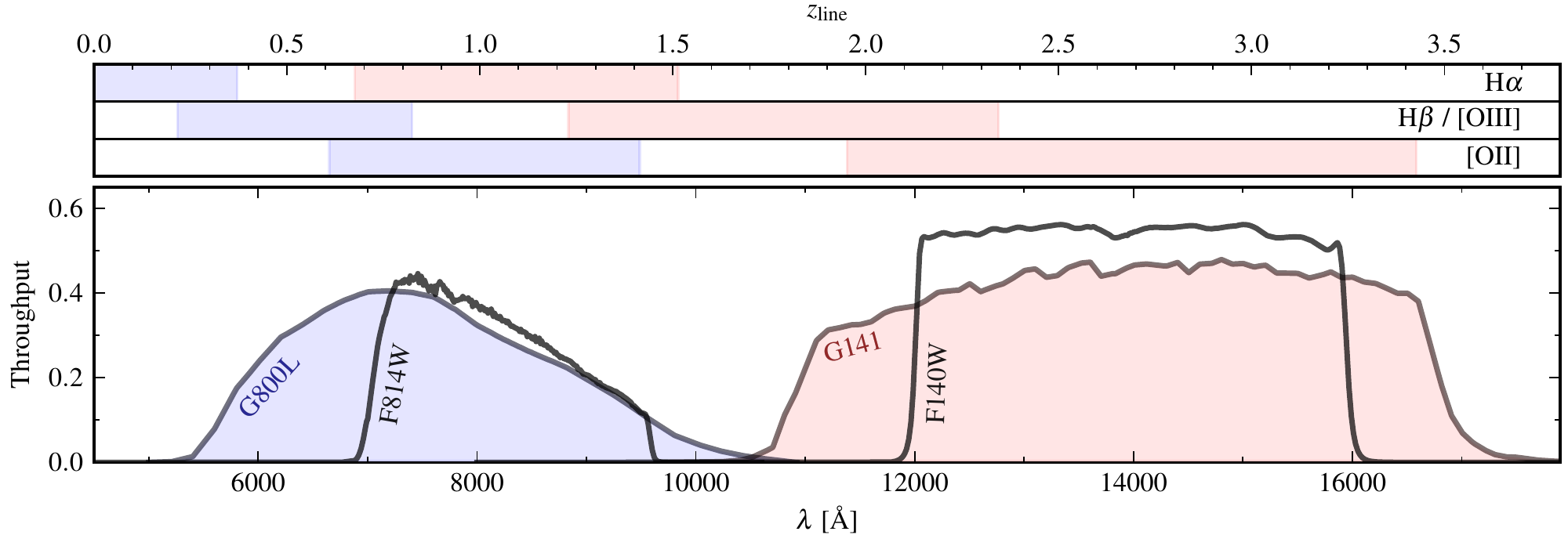}
\caption{Throughput curves of the WFC3/G141 (primary) and ACS/G800L (parallel) grisms and the WFC3/F140W and ACS/F814W imaging filters used to define the wavelength reference for the grisms.  The shaded bands in the top panel indicate the redshift range in which rest-frame optical spectral lines of \ion{O}{2}, H$\beta$, \ion{O}{3}, and H$\alpha$ fall within the coverage of the WFC3 (red) and ACS (blue) grisms.    \label{f:throughput}}	
\end{figure*}

The WFC3 G141 grism is the primary spectral element used for the 3D-HST survey.  The combined transmission of the \HST\ optical telescope assembly and the primary spectral order of the G141 grism (``$+1^\mathrm{st}$'') is greater than 30\% from 1.10 to 1.65 \micron, reaching a peak of nearly 50\% at 1.45 \micron.  The mean dispersion of the $+1^\mathrm{st}$ order is 46.5~\AA pixel$^{-1}$ ($R\sim 130$) and varies by a few percent across the field of view.    The uncertainties of the wavelength zeropoint and dispersion of the G141 grism are 8\AA\ and 0.06\AA pixel$^{-1}$, respectively.  For a full description of the calibration of the WFC3/G141 grism, see \cite{kuntschner:10}.  Spectral features covered by the G141 grism include H$\alpha$ at $0.7 < z < 1.5$, [\ion{O}{3}]$\lambda$5007 at $1.2 < z < 2.3$, [\ion{O}{2}]$\lambda$3727 at $2.0 < z < 3.4$, and the Balmer / 4000~\AA\ break at $1.8 < z < 3.1$ (Figure \ref{f:throughput}).  The nominal G141 dispersion corresponds to $\sim$1000 \kms for H$\alpha$ at $z>1$; however, in practice the resolution of the slitless grism spectra is determined by the physical extent of a given object (see Section \ref{s:redshift_fits}).

Observations with the \HST\ grisms typically require an accompanying image taken with an imaging filter to establish the wavelength zeropoint of the spectra \citep[see, e.g.,][]{kummel:09}.  For 3D-HST, we obtain these so-called ``direct'' images in the broad F140W filter that spans the gap between the standard $J$ and $H$ passbands and lies roughly in the center of the G141 sensitivity (Figure \ref{f:throughput}).  While CANDELS will eventually provide significantly deeper imaging of the 3D-HST fields in the F125W and F160W WFC3 filters, the 3D-HST F140W direct images can be useful for scientific analysis in addition to calibrating the grism, as they reach depths competitive with even the deepest ground-based surveys ($H\sim26.1$, 5$\sigma$) with spatial resolution $\sim0\farcs13$ (see Section \ref{s:image_prep} and also \citealp{vandokkum:11}).

In addition to the primary WFC3 observations, 3D-HST obtains parallel ACS F814W imaging and G800L grism spectroscopy.  The G800L grism covers wavelengths 0.55--1.0 \micron\ with a dispersion of 40~\AA pixel$^{-1}$ and a resolution of 80~\AA\ for point-like sources \citep[][]{kummel:11a}.  The parallel spectroscopy extends the H$\alpha$ line sensitivity of the survey to $z=0$ and will provide coverage of additional lines at certain redshift intervals where only a single line is visible in WFC3/G141, for example [\ion{O}{3}] in G800L and H$\alpha$ in G141 at $0.7 < z < 0.8$ (Figure \ref{f:throughput}).

\subsection{Survey fields}\label{s:survey_fields}

\begin{figure*}
\epsscale{1.1}
\plotone{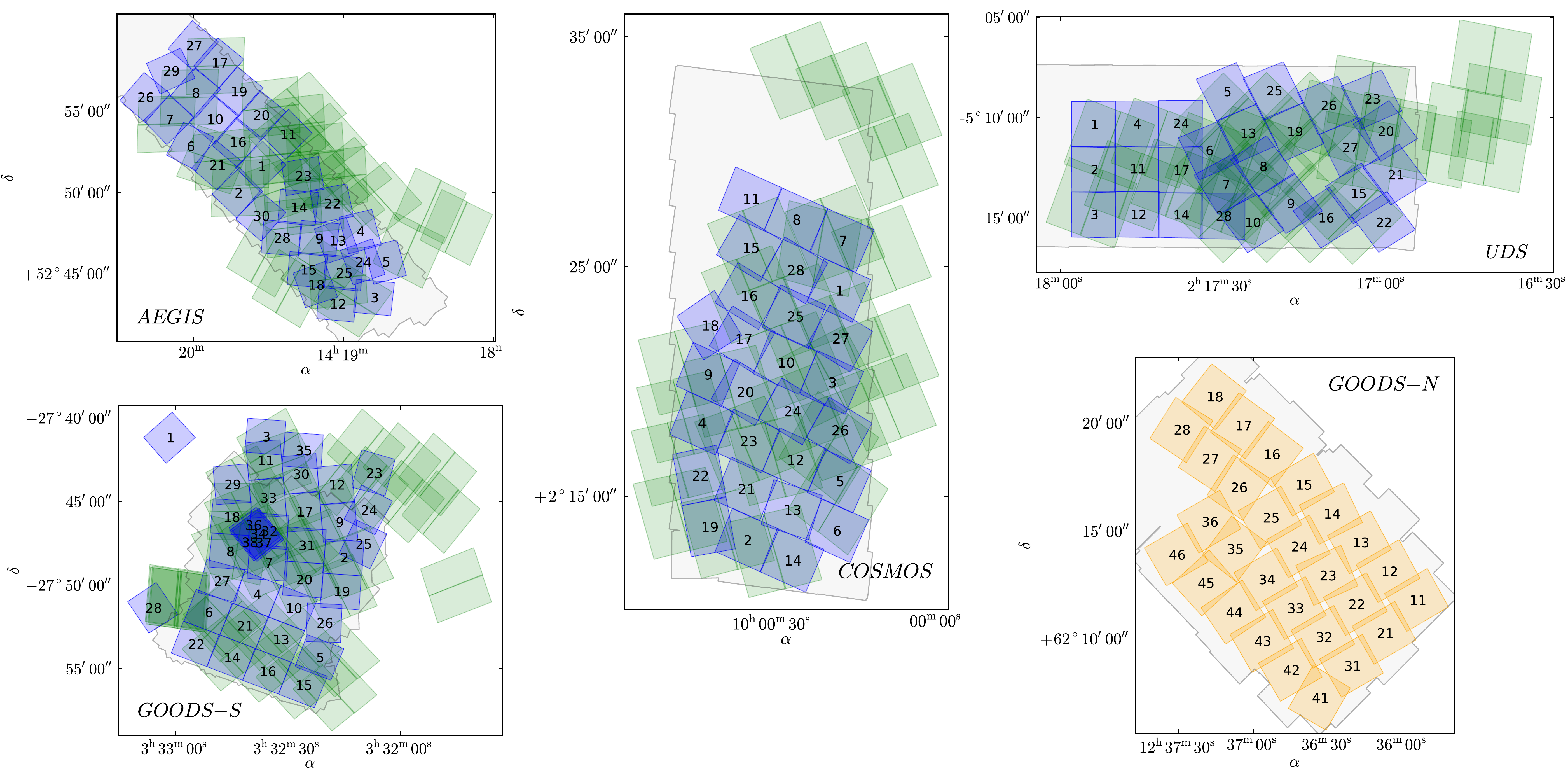}
\caption{Layout of the 124 3D-HST pointings.  Primary WFC3 F140W+G141 pointings are shown in blue with the pointing ID numbers as defined in the \HST Phase II file.  The locations of the parallel ACS F814W+G800L observations are shown in light green.  Also indicated is the distribution of the 28 pointings covering the GOODS-North field from program GO-11600 that are incorporated into 3D-HST.  The light gray polygons indicate the footprint of the CANDELS WFC3 imaging, including both the ``wide'' and ``deep'' components of that survey.  Note that the relative sizes of the separate fields are not shown exactly to scale. \label{f:pointings}}	
\end{figure*}

The 3D-HST survey is allocated 248 primary+parallel orbits over Cycles 18 and 19.  3D-HST will cover roughly 75\% of the area imaged by the CANDELS survey in the EGS/AEGIS, COSMOS, UKIDSS-UDS, and GOODS-South fields (Figure \ref{f:pointings}).  Furthermore, G141 grism coverage of most of the GOODS-N field from program GO-11600 (PI: B. Weiner) is incorporated into 3D-HST as the observational strategy of the GOODS-N observations are nearly identical to that of 3D-HST.  These parts of the sky are the best-studied extragalactic 
survey fields, offering a wealth of deep, multi-wavelength imaging and spectroscopic observations from large number of previous ground-based and space-based surveys.  
\cite{grogin:11} provide a detailed summary of the ancillary datasets available in these survey fields, from X-ray to radio wavelengths, which are crucial for the interpretation of both the WFC3 imaging and grism spectroscopy datasets (see also Section \ref{s:redshift_fits}).

\subsection{Observational strategy and mosaic layout}\label{s:layout}

The 248 3D-HST orbits are divided among 124 individual visits of two orbits each.  The layout of the 3D-HST pointings is shown in Figure \ref{f:pointings} and summarized in Table \ref{t:fields}.  In order to schedule 3D-HST concurrently with CANDELS observations of the same fields over Cycles 18 and 19, initially no ORIENT constraints were imposed on any of the visits.  After scheduling the ORIENTs were fixed and the positions of the individual pointings were optimized to provide contiguous mosaics and maximum overlap between the primary WFC3 G141 and parallel ACS G800L observations, as shown in Figure \ref{f:pointings}.  Owing to this optimization, fully 90\% of the G141 mosaic (excluding GOODS-N) will be covered by between two and four orbits of the ACS grism.  Within the GOODS-South mosaic, four additional two-orbit visits at the same orientation are centered on the Ultra Deep Field (UDF).  The GOODS-South pointings outside of the area with CANDELS coverage provide WFC3 grism spectroscopy of the HUDF09\footnote{\scriptsize{\url{http://archive.stsci.edu/prepds/hudf09/}}} and WFC3-ERS fields \citep[see][]{bouwens:10b}.

The first 3D-HST exposures were obtained in 2010 October and the survey is nearly completed as of 2012 March: all pointings in the COSMOS, GOODS-S, and UDS fields have been observed, and two failed pointings in AEGIS will be re-observed by the end of 2012.  The 28 grism pointings in GOODS-N were completed in 2011 April.\footnote{The fitting analysis and specific spectra described below come from 70 G141 pointings available as of August 1, 2011.}  Due to scheduling constraints, a given field will not have both complete CANDELS and 3D-HST coverage before both surveys are finished.  For example, the two epochs of CANDELS observations in the UDS were completed in 2011 January, while the 3D-HST coverage of that field was only completed in 2012 February.

Each of the 3D-HST two-orbit visits with WFC3 is structured in an identical fashion:  four pairs of a short F140W direct image followed by a longer G141 grism exposure.  The four pairs of direct+grism exposures are separated by small telescope offsets to enable the rejection of hot pixels and pixels affected by cosmic-rays, as well as dithering over some WFC3 cosmetic defects such as the ``IR-blobs'' \citep{pirzkal:10}.  The dither pattern is shown schematically in Figure \ref{f:dither}.  The sub-pixel offsets of the dither pattern are chosen to improve sampling of the WFC3 PSF, which enables some recovery of the image quality lost by the pixels that undersample the PSF by a factor of 2 \citep[see, e.g.,][]{fruchter:02, koekemoer:11}.  Due to the offset and rotation of ACS with respect to WFC3, the pixel subsampling of the ACS parallel exposures is somewhat less than optimal (Figure \ref{f:dither}).

\tabletypesize{\scriptsize}	
\begin{deluxetable}{clrrrc}
\tablecolumns{13} 
\tablewidth{0pt} 
\tablecaption{Summary of the 3D-HST survey fields}
\tablehead{\colhead{Field}          &
		   \colhead{Prog. ID} &
		   \colhead{R.A.}       &
		   \colhead{Dec.}       &
		   \colhead{$N\tablenotemark{a}$}  &
		   \colhead{Sky (\epers)}}
\startdata
AEGIS 		& 12177 & $14:19:31$ & $+52:51:00$ & 30  & 0.9 \\
COSMOS 		& 12328 & $10:00:29$ & $+02:20:36$ & 28 & 2.4 \\
GOODS-South & 12177 & $03:32:31$ & $-27:48:54$ & 32 & 1.4 \\
HUDF09 		& 12177 & $03:32:39$ & $-27:47:01$ & 3\tablenotemark{b} & 1.2 \\
UKIDSS-UDS 	& 12328 & $02:17:26$ & $-05:12:13$ & 28 & 1.4 \\
GOODS-North & 11600\tablenotemark{c} & $12:36:50$ & $+62:14:07$ & 28 & 1.0 \\
\enddata
\tablenotetext{a}{$N$ is the number of independent WFC3 pointings within the survey fields.}
\tablenotetext{b}{The main HUDF09 field is covered by four visits.  The flanking HUDF09-1/2 fields have one visit each.}
\tablenotetext{c}{Incorporated into 3D-HST; PI: B. Weiner}
\label{t:fields}
\end{deluxetable}


All of the WFC3/IR exposures are obtained in MULTIACCUM mode using either a SPARS50 (F140W) or SPARS100 (G141) read-out sequence.  All of the direct exposures have NSAMP=5, corresponding to individual exposure times of 203 s and a total of 812 s for each visit/pointing.  Depending on the varying usable length of an orbit, the G141 exposures have NSAMP=12--15 and the total grism exposure times range from 4511--5111 s per visit/pointing.  The ACS F814W direct images all have total exposure times of 480 s per visit, and the G800L grism images have exposure times ranging from 2925 to 3523 s.  The ACS exposure times are somewhat less than those of WFC3 as a result of the larger overheads for reading out the ACS/WFC detectors.

%
%
\section{Data reduction}\label{s:data}

\subsection{Pre-processing}\label{s:preprocessing}
We use as a starting point the standard calibrated data products (images with the \textsc{flt} extension) provided by the the \HST archive that have been processed by the \texttt{calwf3} and \texttt{calacs} reduction pipelines for the WFC3 and ACS 3D-HST exposures, respectively.  The calibrated WFC3/IR images have 1014$\times$1014 pixels, with roughly $0\farcs128$ pixel$^{-1}$.  The ACS/WFC images consist of two 4096$\times$2048 pixel extensions with roughly $0\farcs05$ pixel$^{-1}$.  Briefly, the calibration pipelines flag known bad pixels in the data quality image extensions, subtract the bias structure determined from zeroth reads and overscan regions, subtract dark current, and apply multiplicative corrections for the detector gain.  Additionally, the pipelines apply a multiplicative flat-field correction to the WFC3/F140W and ACS/F814W direct images.  The flat-fielding of the grism exposures is discussed below in Section \ref{s:background_subtraction}.  Finally, bias striping \citep{grogin:10} and charge-transfer efficiency (CTE) \citep{anderson:10} corrections are applied to the ACS \textsc{flt} pipeline products.  A more comprehensive description of the WFC3 and ACS reduction pipelines is given by \cite{koekemoer:11}.  

\subsection{Image preparation}\label{s:image_prep}

\begin{figure}
\epsscale{1.}
\plotone{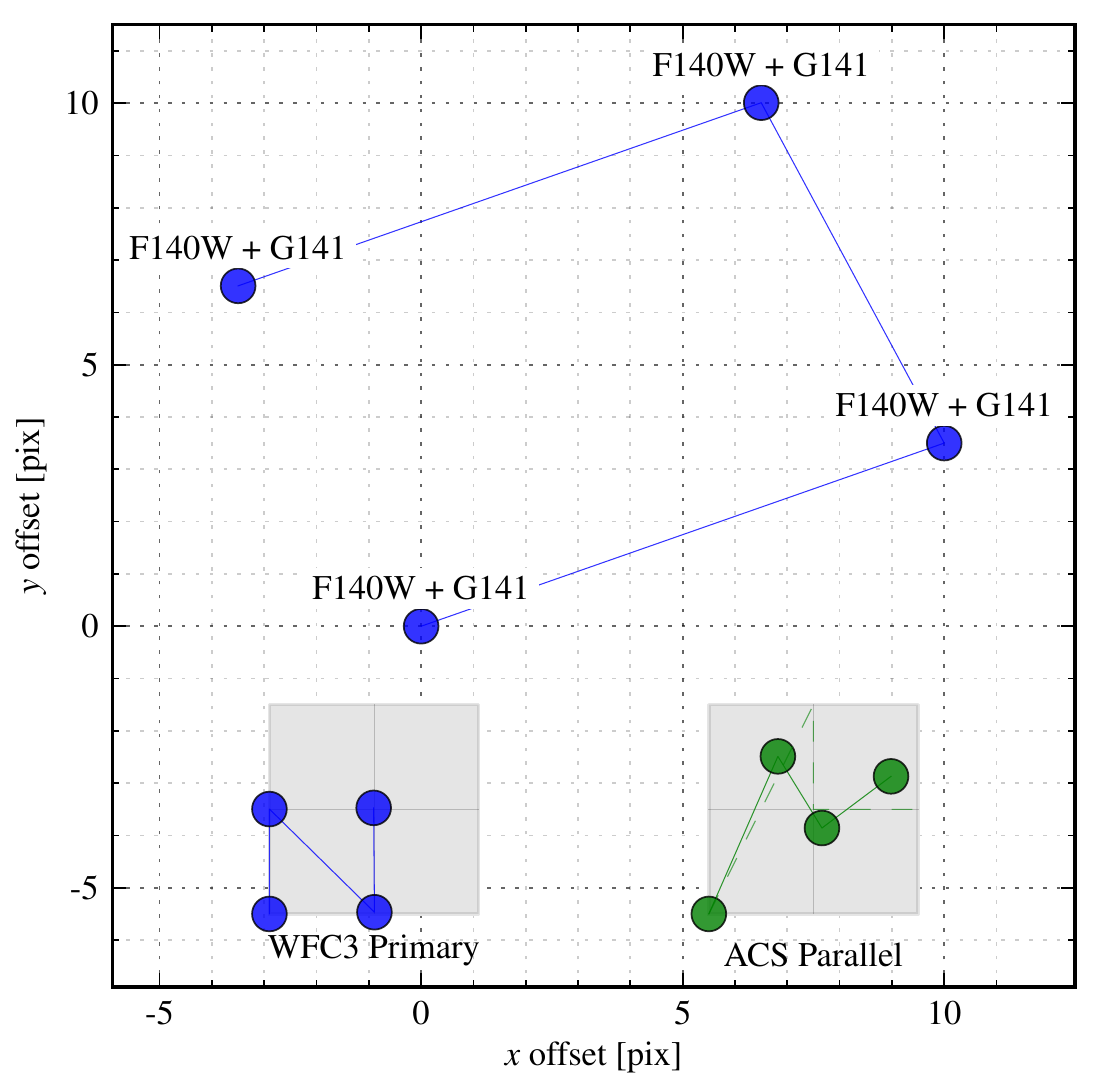}
\caption{Dither pattern in WFC3 pixels for each two-orbit visit, starting at the origin (0,0).  At each position, a short F140W direct exposure is followed by a longer G141 exposure at the same position.  The insets show the sub-pixel sampling of the dither pattern for the four primary WFC3 (blue) and four parallel ACS (green) exposures.  The WFC3 pixel is fully sampled at half-pixel intervals, potentially allowing optimal image combination with interpolation.  The sampling of the parallel ACS pixels (solid line) is similar, but not identical, to that of optimal sampling (dashed line).  \label{f:dither}}	
\end{figure}

With the calibrated images obtained as described above, we perform a number of additional preparation steps in order to produce mosaics of each visit independently, which are composed of the four-exposure sequences described in Section \ref{s:layout}.  First we use the \texttt{MultiDrizzle} software \citep{koekemoer:02} to identify any hot pixels or cosmic rays not flagged by the instrument calibration pipelines.  We use essentially the same strategy as described by \cite{koekemoer:11}, though we note that we had to increase the first-pass detection threshold of the cosmic ray rejection to avoid rejecting the central pixels of stars.  The final data product after applying the alignment and sky-subtraction steps described below is an undistorted mosaic with a final scale of $0\farcs06$ pixel$^{-1}$ (with \textsc{PIXFRAC}=0.8), which is justified by the optimal sampling of the WFC3 PSF by the 3D-HST dither pattern (Figure \ref{f:dither}).  The 5$\sigma$ depth of the F140W detection images is $H_{140}\approx26.1$ for point sources within a $0\farcs5$ diameter aperture, varying by $\pm$0.3 mag due to the field-dependent background levels (see Table \ref{t:fields}).

\subsubsection{WCS alignment}

Small adjustments to the commanded telescope dither offsets are determined using the PyRAF routine, \texttt{tweakshifts}.  These shifts are at most 0.1 pixels.  We refine the WCS coordinates of the mosaic image with respect to a WCS reference image by matching object catalogs extracted from each image and fitting for shifts and rotations using the IRAF task, \texttt{geomap}.  The adopted WCS reference images are the following \HST-based public image mosaics: AEGIS, ACS-$i_{814}$ \citep{davis:07}; COSMOS, ACS-$i_{814}$ \citep{koekemoer:07}; GOODS-N, ACS-$z_{850}$ \citep{giavalisco:04}; GOODS-South, ACS-$z_{850}$ \citep{giavalisco:04}; HUDF09, WFC3-$H_{160}$ \citep{bouwens:10b}; and UDS, WFC3-$H_{160}$ \citep{koekemoer:11}.  The public CANDELS mosaics are used when available, and future CANDELS data releases will be adopted for those fields currently using ACS data products.  The derived image rotations are typically less than 0.1 deg, and the rms of the shifts matched between the catalogs is generally $\sim$0.1 pixel ($0\farcs006$).  Once the relative and offset shifts are determined for the direct images, the grism exposures are assigned the same shift as their preceding direct image, assuming that there was no shift between the two exposures (as commanded).

\subsubsection{Background subtraction and grism flat-fielding}\label{s:background_subtraction}

\begin{figure}
\epsscale{1.}
\plotone{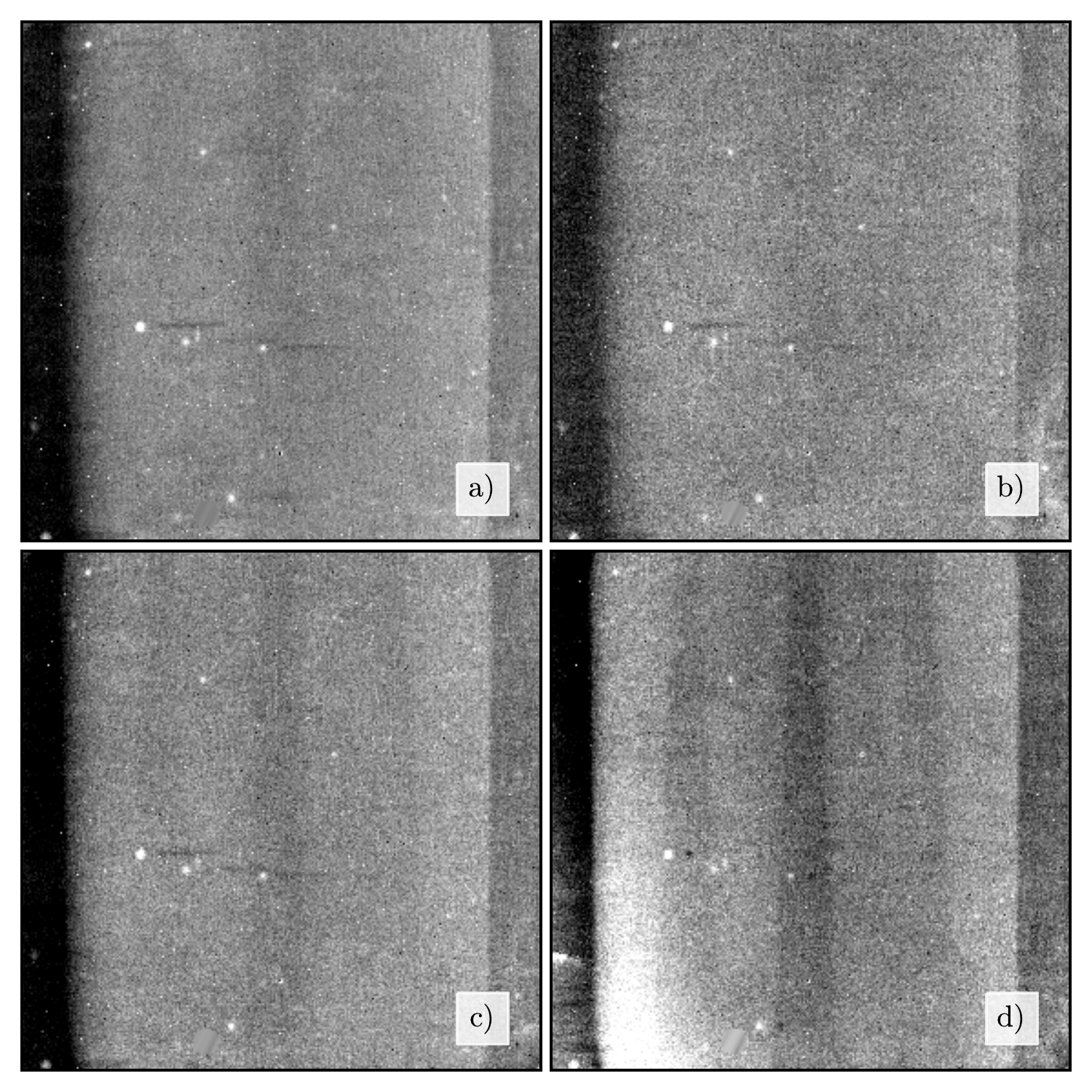}
\caption{Master background images for the G141 grism.  Each master image is a masked median combination of 10 or more individual \textsc{flt} images that show similar structure in the background, such as the relative contrast of the dark strips on either side of the image.  The F140W flat-field was divided out before the image combination.  Black/white parts of the displayed images correspond to deviations of $\pm$~5\%.
\textbf{a)} All available COSMOS pointings, \textbf{b)} GOODS-N ``low'', \textbf{c)} GOODS-N ``high'', and \textbf{d)} GOODS-N ``very-high''.  The structure in the GOODS-N background images correlates roughly, but not exactly, with the overall background level.  Using only a single master background image, such as the one provided with the aXe calibration files, can result in significant amount of residual structure in the background-subtracted images. \label{f:background_images}}	
\end{figure}

While the near-IR background is much lower in low-Earth orbit than it is from the ground, the background flux is still a significant component that must be subtracted from both the direct and grism exposures.  Subtracting the background from the direct images is done in the following way: we subtract a second-order polynomial fit to each exposure after aggressively masking objects detected in the \texttt{MultiDrizzle} mosaic and mapped back to the distorted frame using the PyRAF \texttt{blot} routine.  A polynomial fit with order greater than zero is necessary as some exposures show structure in the background likely caused by enhanced airglow for a particular pointing orientation with respect to the Earth limb.

Subtracting the background from the grism exposures is more difficult than for the direct images.  There is significant structure in the grism background that is a result of a superposition of multiple dispersed grism orders and the fact that the finite field of view of the entrance window causes different regions of the detector to see different combinations of the spectral orders.  Effort has been made to produce a ``master'' grism background image created from the (masked) average of many grism exposures \citep{kummel:11b}, which can then be scaled to and subtracted from a particular grism exposure.  This is the approach followed by the \aXe\ software package \citep{kummel:09}.

\begin{figure}
\epsscale{1.15}
\plotone{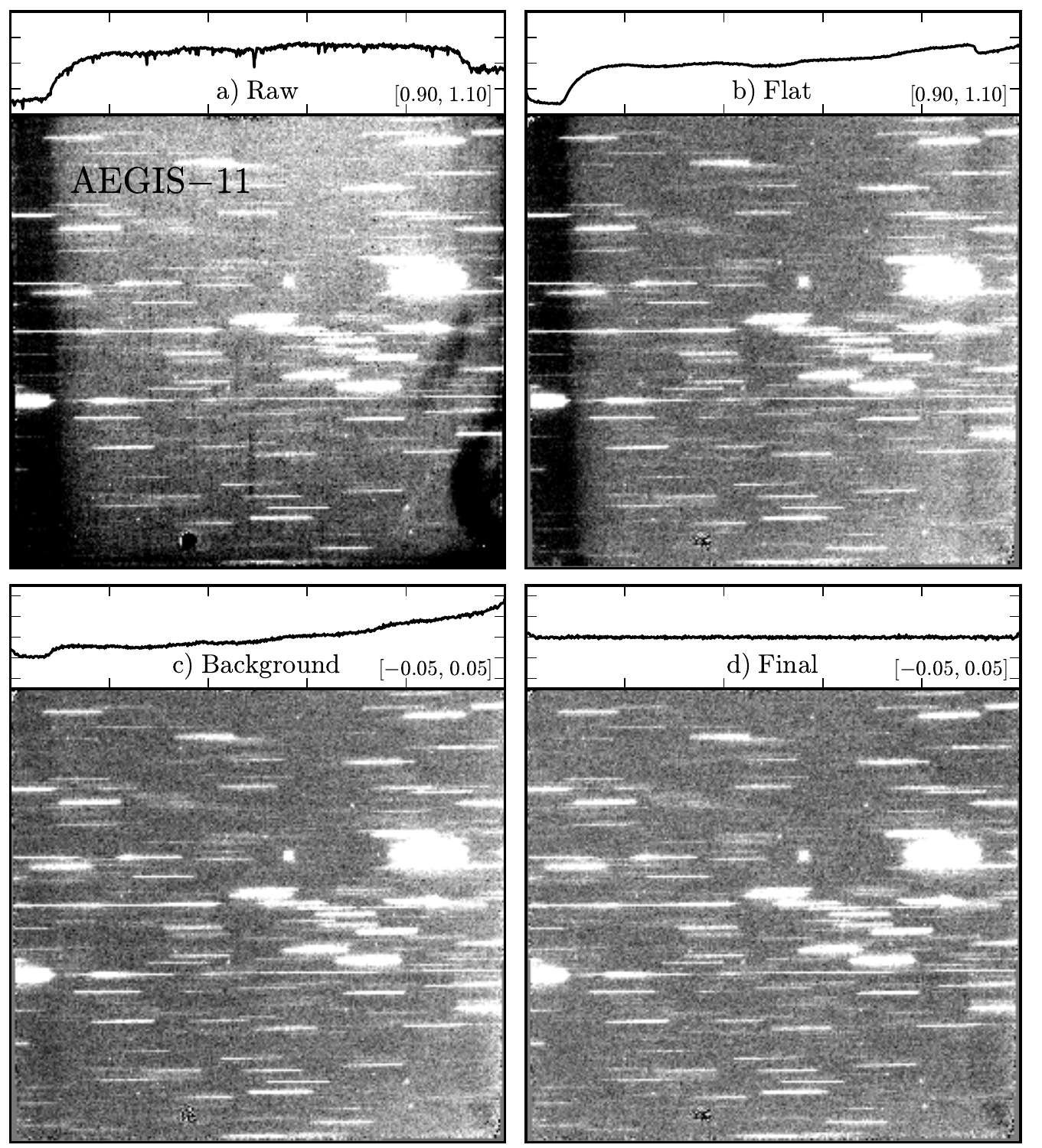}
\caption{Flat-field correction and background subtraction for a single 3D-HST pointing (AEGIS-11). \textbf{a)} A combination of the raw grism images shows the combined features of the background and the WFC3 flat-field features such as the ``wagon wheel'' at lower right. \textbf{b)} Dividing by the F140W imaging flat removes the flat-field features.  The wavelength-dependent component of the flat-field is generally less than 2\% (Appendix \ref{ap:flat_field}), and applying a single flat-field greatly simplifies the data analysis. \textbf{c)} Subtracting by the best (scaled) master background image (see Figure \ref{f:background_images}) removes most of the remaining structure in the background.  \textbf{d)} The low-level structure that remains after subtracting the master background is largely parallel to the image columns and is removed by subtracting the average along columns after all object flux has been aggressively masked.  The top panels show the image counts averaged along columns with the full vertical plot range indicated by the numbers in brackets (in $\mathrm{e}^{-} /\mathrm{s}$).  For reference, a compact object with $H_{140}=23$ has a peak flux of $\sim$~0.05~$\mathrm{e}^{-} /\mathrm{s}$.\label{f:subtract_background}}	
\end{figure}

The 3D-HST fields span a wide range of celestial coordinates that result in distinct orientations of the instrument with respect to the zodiacal and Earth glow sources of the background light.  From the large number of 3D-HST grism observations that have already been obtained, we find that a single master G141 background image is insufficient to explain the observed variety of structure in the background.  We therefore correct the background in the following way.  We produced four representative master background images that are median (masked) combinations of subsets of 3D-HST grism exposures that have similar overall structure, determined by eye.  These images are shown in Figure \ref{f:background_images}.  The primary features in the grism background are the dark vertical bands spanning roughly 100 pixels at the left and right sides of the image.  The relative intensity of these bands and the sharpness of their edge is correlated with the overall background level, but they can also vary at a given background level for images in different survey fields.    There are dark horizontal bands in some of the master images that look like negative grism spectra.  These features are the result of the grism ``dispersing'' the IR-blob features of decreased sensitivity, which are not physically located on the detector itself but rather on the Channel Select Mechanism in the WFC3/IR optical path \citep{pirzkal:10}.  

\begin{figure*}
\epsscale{1.}
\plotone{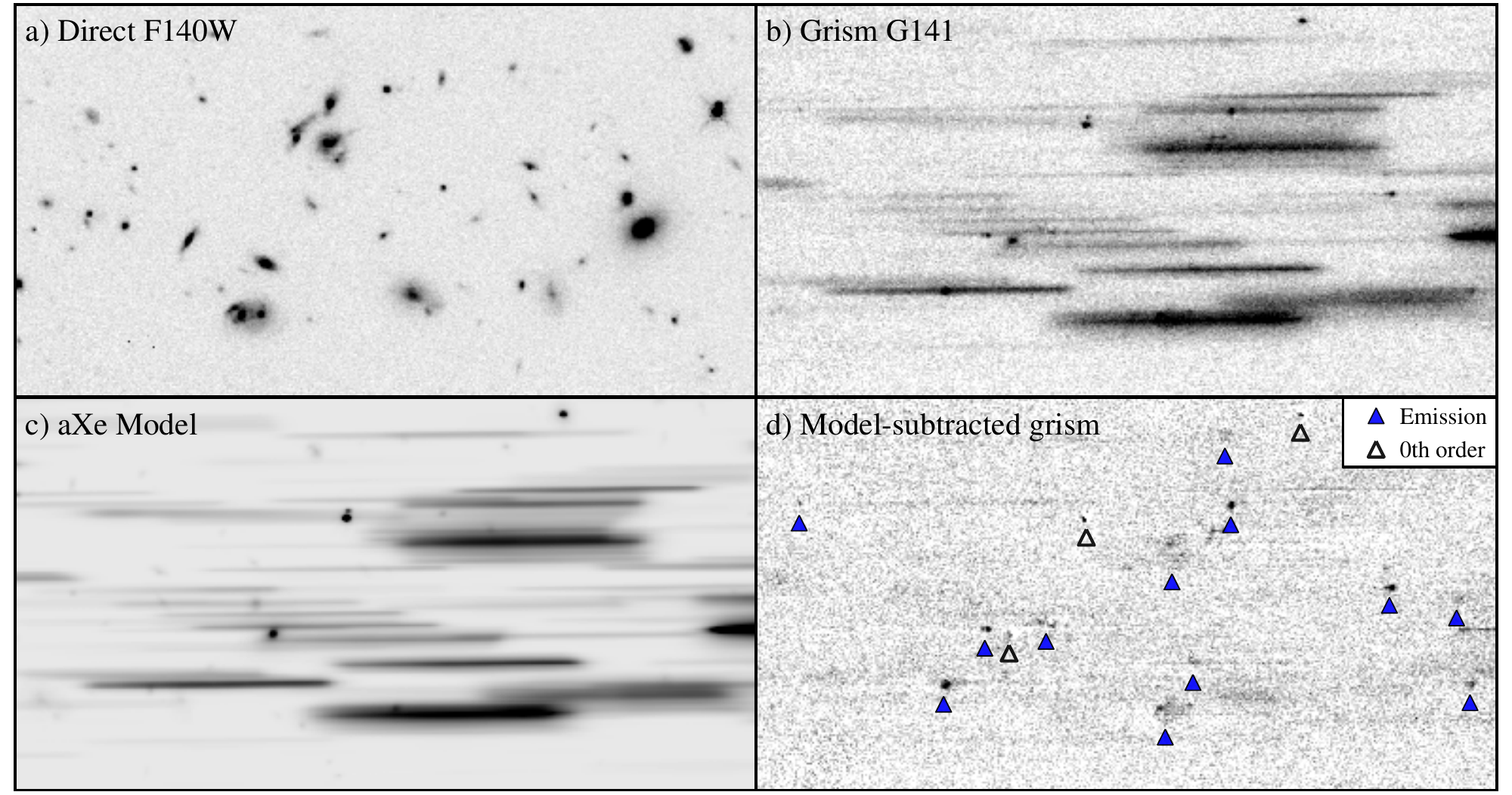}
\caption{Model of the grism spectra based on the observed direct images and computed with \aXe.  Panels \textbf{a} and \textbf{b} show $50\arcsec\times28\arcsec$ cutouts of the F140W and G141 observations within the GOODS-South field, with wavelength increasing towards the right on the grism panel.  Panel \textbf{c} shows the \aXe\ ``fluxcube'' model of the grism spectra, where the spatial profile and intensity of the spectra are determined from the direct image(s).  Along with the F140W imaging, the model for the example shown includes color information from the F125W and F160W CANDELS imaging of this field.  Panel \textbf{d} shows the model-subtracted grism image, with the image stretch increased by a factor of two compared to panels \textbf{b} and \textbf{c}.   The model is quite a good representation of the data, despite the fact that no fit has been done; the model inputs come from the direct image(s) and the grism calibration alone. Compact features in the model are zeroth order spectra.  While the zeroth order spectra are not perfectly modeled and subtracted (open triangles in panel \textbf{d}), they can generally be identified and distinguished from emission lines.  The majority of the residuals in panel \textbf{d} are emission lines, indicated by filled blue triangles.  Even this small cutout shows the diversity of the emission lines found within the 3D-HST survey, which is fully $\sim$1600 times larger than the area shown here.  \label{f:grism_model}}	
\end{figure*}

The master background images shown in Figure \ref{f:background_images} look very different from the master background image produced by \cite{kummel:11b} and distributed with the G141 calibration files.  For the images in Figure \ref{f:background_images} we have divided out the F140W imaging flat-field \textit{before} creating the image combinations in order to separate the pixel-to-pixel variations of the flat-field from the more smoothly-varying background.  We find that the wavelength dependence of the flat-field is less than $\sim$1\% across the G141 sensitivity (Appendix \ref{ap:flat_field}).  This is similar to the uncertainties of the overall G141 flux calibration, and we therefore opt for a simplified approach by dividing out the single F140W flat-field from all of the G141 grism exposures.  We then subtract the scaled master background image whose structure best matches that of a particular exposure, as determined from a $\chi^2$ test.  Note that it is possible that two subsequent grism exposures within a given visit require different master background images depending on the varying background within an orbit (Appendix \ref{ap:background_variation}).  Finally we subtract the median level of the resulting (masked) background pixels averaged along image columns.  While adopting the variable background images greatly improves the overall background subtraction compared to using the single image, we find that this last step is necessary to remove any residual structure in the background, which tends to be parallel to image columns.  The complete process of first dividing by the flat field, then subtracting the master background image, and finally removing the $x$-dependent residuals is demonstrated in Figure \ref{f:subtract_background}.  The background in the final corrected grism images is typically flat to better than 1\%.

We have described the flat-fielding and background subtraction in considerable detail as these steps are crucial for minimizing systematic effects in the spectra for this blank-field spectroscopic survey.  Local background subtraction is not feasible for the relatively deep 3D-HST grism exposures because it is generally impossible to identify pixels adjacent to a given object that will be entirely free of flux contributed by nearby overlapping spectra.  Small uncorrected errors in the background subtraction can later be manifested as continuum break features in the extracted spectra, particularly at fainter magnitudes where 3D-HST can provide truly unique near-IR spectra currently unobtainable from the ground.

\subsection{Extraction of the grism spectra}\label{s:spectra_extraction}

We use the \aXe\ software package \citep{kummel:09} to extract the grism spectra in much the same way as described by \cite{vandokkum:10a} and \cite{atek:10}.  The primary inputs to the extraction software are a detection image mosaic (F140W or F814W) and the individual background-subtracted grism exposures generated as described in Section \ref{s:image_prep}.  An object catalog is generated from the detection image with the \texttt{SExtractor} software \citep{bertin:96}, which also produces a segmentation map that indicates which pixels in the direct image are assigned to each object.  For a given pixel within a particular object's segmentation map, the calibration of the \HST\ grisms determines where the dispersed light from that pixel will fall on the grism exposures, with the pixel in the direct image defining the wavelength zeropoint for the spectrum.  Thus, the grism spectrum of an object is the sum of all of the dispersed pixels within that object's segmentation image, or rather, the spectrum is a superposition of the object profile at different wavelengths offset by the grism dispersion.  The effective spectral resolution is therefore a combination of the grism dispersion and the object profile in the sense that the effective resolution decreases with increasing object size in the dispersion direction (see also Section \ref{s:redshift_fits} and Figure \ref{f:example_spectra}).

Because no slit mask is used and the length of the dispersed spectra is larger than the average separation of galaxies down to the detection limit of the 3D-HST survey, the spectra of nearby objects can overlap.  This ``contamination'' of an object's spectrum by flux from its neighbors must be carefully accounted for in the analysis of the grism spectra.  We use \aXe\ to produce a full quantitative model of the grism exposures using the information in the direct image as described above.  This is the \aXe\ ``fluxcube'' contamination model, which makes use of the spatial information contained in the high-resolution \HST\ images to model the two-dimensional (2D) grism spectrum.  In order to generate a model spectrum based on the direct image, we take the observed F140W flux and full spatial profile within a given object's segmentation map and assume a constant profile and a flat spectrum in units of $f_\lambda$.  As additional \HST\ photometric bands become available, for example the CANDELS F125W and F160W imaging covering the 3D-HST survey fields, they will be added to the \aXe\ fluxcube to incorporate the wavelength dependence of both the flux (i.e., the color) and spatial profile into the model.  

The relationship between the direct image and the grism spectra and a demonstration of the fluxcube spectral model are shown in Figure \ref{f:grism_model}.  The images are oriented as in the individual \textsc{flt} exposures, with the grism spectra offset in the positive \textit{x} direction with respect to the direct image and with wavelength increasing towards the right.  The main horizontal features in Figure \ref{f:grism_model}b are the ``$+1^\mathrm{st}$'' spectral order, which has the greatest sensitivity \citep{kuntschner:10}.  Compact, point-like features seen in both the observed and model images are $0^\mathrm{th}$ order spectra.  The residuals of the model subtracted from the grism image are shown in Figure \ref{f:grism_model}d.  The model is generally a reliable quantitative representation of the data.  It bears mention that for the present reduction there has been no \textit{fit} to optimize the model---the relatively low level of the residuals demonstrates the quality and stability of the G141 grism calibration.  Most of the compact features in the residuals are in fact emission lines, which are not included in the model.  The $0^\mathrm{th}$-order spectra do not always subtract completely, but their presence in the model allows them to be distinguished from emission lines.  In future versions of the reduction, we will implement an iterative scheme to refine the spectral model based on the observed spectra.

We use \texttt{aXeDrizzle} to combine the four grism exposures (in the original distorted \textsc{flt} frame) of each visit/pointing and extract a 2D spectrum for each object  with perpendicular spatial and dispersion axes.  We adopt output $0\farcs06\times22$~\AA\ spectral pixels, which are roughly square with respect to the drizzled pixels in the direct image mosaics.  These 2D spectra with \HST\ spatial resolution ($\sim0\farcs13$ for WFC3/G141) are one of the truly unique products of the 3D-HST survey.  As there is no slit defining a spatial axis, an ``effective slit'' running roughly parallel to the major axis of each object is not generally parallel to the \textit{y} pixel direction in the undistorted frame.  The \aXe\ software includes an option to account for the orientation of this effective slit to optimize the wavelength resolution of the extracted spectra \citep[see][]{kummel:09}; however, we simply adopt the spatial axis as perpendicular to the spectral trace and account for the orientation and shape of the object profiles in post-processing analysis (see Section \ref{s:redshift_fits}).  Finally, we use \aXe\ to extract optimally-weighted 1D spectra from the drizzled 2D spectra, where the relative weights are determined from the object profile perpendicular to the dispersion axis.  

\subsection{Grism sensitivity}\label{s:grism_sensitivity}

We have described above how the non-negligible background emission must be subtracted from the grism exposures.  In fact, this background flux from a combination of zodiacal light, Earth glow, and low-level thermal emission is the limiting factor for the sensitivity of the 3D-HST grism exposures: the read noise of the WFC3/IR detector is $\sim$20 \textit{e}$^{-}$, while the number of background electrons per pixel in a typical grism exposure is 1.4 \epers $\times$ 1300 s = 1820 \textit{e}$^{-}$.  The average background level varies for the different survey fields, as shown in Table \ref{t:fields}, depending on the field location relative to the ecliptic plane and the date of observation.  In general, the observed background levels are consistent with those predicted by the WFC3 Exposure Time Calculator (ETC).  The angle of the bright Earth limb can vary within an orbit, and the two grism exposures taken within an orbit can have background levels that differ by as much as 50\% (Appendix \ref{ap:background_variation}) and also different background structure (Section \ref{s:background_subtraction}).

\begin{figure*}
\epsscale{1.1}
\plotone{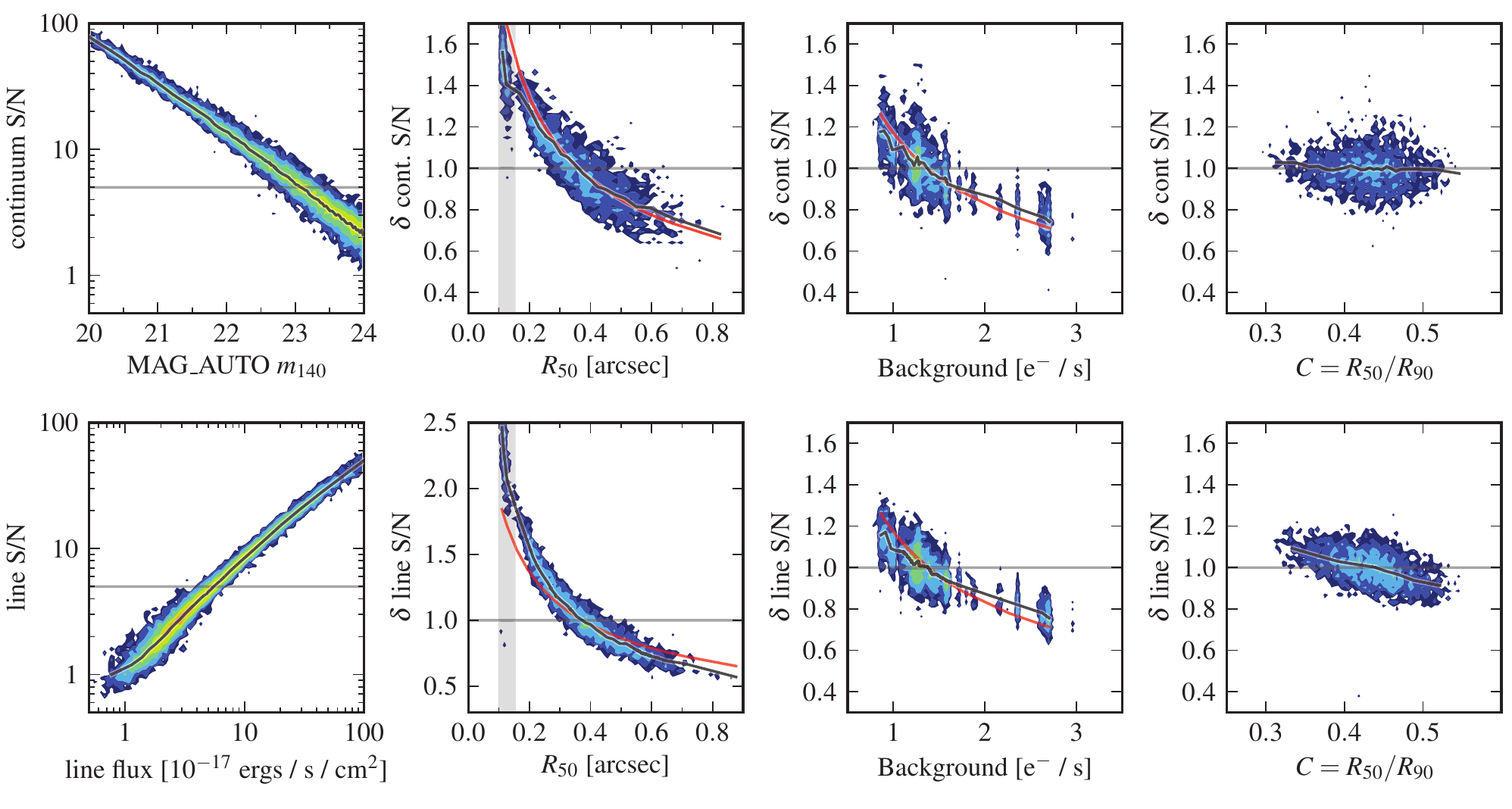}
\caption{Simulations of G141 spectra to evaluate the 3D-HST continuum (top panels) and emission line (bottom panels) sensitivities.  The input spectrum for each of 13~000 simulated galaxies is a flat continuum ($f_\lambda$) and a single (narrow) emission line at 1.3\micront\ with equivalent width 130~\AA.  The left panels show how the continuum (per 92~\AA\ resolution element) and line signal-to-noise varies with continuum magnitude and integrated line flux, respectively. The additional panels show how the S/N depends specifically on higher-order properties of individual galaxies:  half-light radius, background level, and morphological concentration, defined as the ratio of the radii containing 50\% and 90\% of the flux in the direct image (SExtractor \flux_radius; the indicated range of $C$ corresponds roughly to Sersic profiles with $n=4$ and $n=1$ from left to right).  In all of the panels, the S/N dependence on the properties other than the one plotted on the ordinal axis have been divided out, i.e., cuts in a 5-dimensional plane, S/N $= f(S,\ R_{50},\ \mathrm{Bkg.},\ C)$.  The shaded gray bands in the second panels indicate $R_{50}$ typical of point sources for $0\farcs06$ pixels.  The solid red lines in the center two panels indicate a dependence of $\delta\mathrm{S/N} \propto 1/\sqrt{x}$. \label{f:spec_sn}}
\end{figure*}

We evaluate the effective continuum and emission line sensitivities of 3D-HST using a suite of simulations that is tied closely to the observed F140W direct and G141 grism exposures.  We use a custom developed software package modeled closely after the \texttt{aXeSIM} package \citep{kummel:09} to generate a 2D model spectrum based on (1) the spatial distribution of flux as determined in the F140W direct image, and (2) an assumed input (1D) spectrum, normalized to the F140W flux.  The 2D grism spectrum is then determined uniquely by the grism configuration files provided by STScI that specify how the flux in a given pixel of the direct image is dispersed into the spectrum of the grism image.  These scripts will be described in more detail and released to the community in a subsequent publication;  for a simple spectral model of a flat continuum, they produce nearly identical results to \texttt{aXeSIM} and the \texttt{aXe} fluxcube.

The primary advantage of the custom software is that we can easily modify the full input spectrum used to generate the model:  here we assume a simple continuum, flat in units of $f_\lambda$, combined with a single emission line at 1.3\micront\ (i.e., H$\alpha$ at $z=1$).  The emission line has a fixed equivalent width of (arbitrarily) 130~\AA, observed frame, and the overall normalization of the spectrum (and thus the integrated line flux) is set to the \textsc{flux\_auto} flux measured by SExtractor on the F140W image.  The result is very much like the \texttt{aXe} ``fluxcube'' model shown in Figure \ref{f:grism_model}, however, each modeled spectrum has the same line+continuum shape.  We add realistic noise to the simulation using the WFC3/IR noise model in the error extension of the FLT images, which includes terms for the poisson error of the source counts and the read noise and dark current of the WFC3 IR detector.\footnote{The simulations do not, however, account for the effects of drizzling, which tends to smooth out some of the apparent noise because the pixel errors tend to be correlated \citep[e.g.,][]{casertano:00}.} Thus, the simulations fully account for noise variations as a function of background level across all of the available pointings, and, most importantly, for the true distribution of source morphologies as a function of brightness within the 3D-HST survey.  There are approximately 13 000 objects in the simulation.

After computing the full grism image models, we extract individual spectra with the standard optimal extraction weighting \citep{horne:86} and measure the median continuum signal-to-noise (S/N) between 1.4--1.6\micront, averaged over a typical G141 resolution element of 92~\AA.  The emission line strengths are extracted using the technique described in Section \ref{s:redshift_fits} and uncertainties on the line fluxes are determined with an MCMC fit of the line + continuum template combination.  Thus, the continuum and line S/N is measured in a very similar way as is done in the analysis of the observed spectra.  Because of the optimal weighting in the spectral extraction, the effective ``aperture'' of the extraction is likely somewhat larger (and variable) compared to the $1\times3$ pixel extraction window used by the WFC3 ETC.  

The result of these simulations is shown in Figure \ref{f:spec_sn}.  The continuum and emission-line S/N depend, clearly, on the signal itself (S), the morphological properties of the objects and the noise properties of the grism exposures.  We explore the dependence of the grism S/N on the signal itself, the object half-light radius, $R_{50}$ (SExtractor 50\% \flux_radius), the background count rate, and a concentration index defined as $C=R_{50}/R_{90}$.  The dependence of S/N on all parameters but the one shown on the ordinal axis is removed in each panel to isolate the contribution of each individually.  After the strength of the signal itself, the S/N depends strongly on the object size, $R_{50}$, and background level, with only a weak, if any, dependence on concentration.  The brightness where the average curves shown in black in the right panels cross the S/N=5 threshold---$m_{140}=23.1$, $f_\lambda=5.5\times10^{-17}\,\fluxcgs$---represents an \textit{average}, not the absolute, sensitivity of the survey.  The properties of the ``average'' galaxy are indicated by where the curves cross $\delta=1$ in the right three sets of panels in Figure \ref{f:spec_sn}:  ($R_{50}$,~background,~$C$)~$\sim$~($0\farcs36$,~1.4~\epers,~0.44).  For the optimal case of point sources and the minimum background, one can read the S/N scaling directly off the figure panels: the 5$\sigma$ continuum and line limits are approximately $1.4\times1.2=1.7$ and $2\times1.2=2.4$ times fainter, respectively, or $23.7$ mag and $2.3\times10^{-17}~\fluxcgs$.  The sensitivity limits in the high-background COSMOS pointings ($>$2~\epers, Table \ref{t:fields} and Appendix \ref{ap:background_variation}) are brighter than the average limits by a factor of $\sim$1/0.8 (0.24 mag).

\begin{figure}
\epsscale{1.2}
\plotone{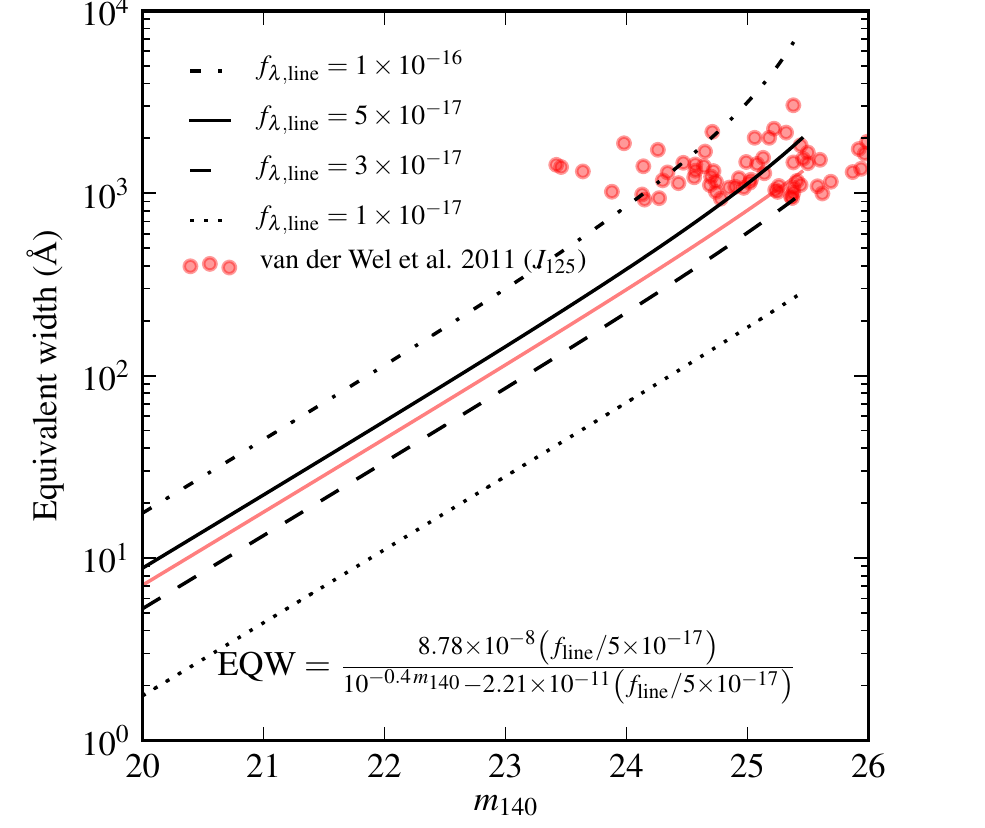}
\caption{Emission line equivalent width (observed frame) as a function of line flux and broad-band F140W magnitude, calculated analytically assuming a spectrum composed of a single emission line and a continuum flat in units of $f_\lambda$.  The analytic function indicated in the inset diverges as the denominator goes to zero, that is, zero continuum and the broad-band magnitude determined by the line flux alone.  Curves are provided for the indicated line fluxes, which have units of $\fluxcgs$.  The high equivalent-width ([\ion{O}{3}]$\lambda5007$) sample of \cite{vanderwel:11} is shown in the red points, where the magnitude was measured in the F125W filter.  The equivalent width sensitivity for F125W (red curve) shifts down by 24\% with respect to F140W, roughly the difference in the filter bandwidths.  The equivalent width sensitivities shown here are uncorrected for underlying absorption, which can be of order 4~\AA\ for the H Balmer lines \citep{savaglio:05}.
\label{f:eqw_mag}}	
\end{figure}

The second column of panels in Figure \ref{f:spec_sn} shows that the continuum S/N is roughly $\propto 1/\sqrt{R_{50}}$ while the S/N of the emission lines goes more like $\propto 1/R_{50}$.  This effect is expected in the sense that the continuum sensitivity only depends on the object extent in the spatial direction as any extent along the spectral axis is averaged out by the dispersion of the grism.  In the case of emission lines, however, it is the pixel area of the emission line extent that determines their S/N as more extended lines are spread out over more noise from the background in both dimensions.  Both the line and continuum S/N roughly follow the square root of the background counts (third column, Figure \ref{f:spec_sn}).  The result that the observed trends are slightly flatter than $1/\sqrt{x}$, where appropriate, likely reflects additional higher-order effects that affect the S/N on the level of $\sim$5\%, whose full characterization is beyond the scope of the present work.

Given the limiting line flux determined as in Figure \ref{f:spec_sn}, one can calculate analytically the limiting line equivalent width as a function of observed broad-band magnitude, where the broad-band passband contains the line and assumptions are made about the continuum shape.  Figure \ref{f:eqw_mag} shows this limiting equivalent width as a function of magnitude in the F140W filter computed for different integrated line fluxes assuming a flat continuum ($f_\lambda$).  We note that the input equivalent width assumed in the simulations described above does not figure into this calculation.  At the faint end, $m_{140}>24$, 3D-HST is sensitive to (observed frame) equivalent widths $\gtrsim500$~\AA.  At the bright end, the equivalent width limit will reach a floor when the poisson noise of the continuum flux is similar to the background noise, which occurs only at $m_{140} \lesssim 19$.

\subsection{Number counts}

\begin{figure}
\epsscale{1.2}
\plotone{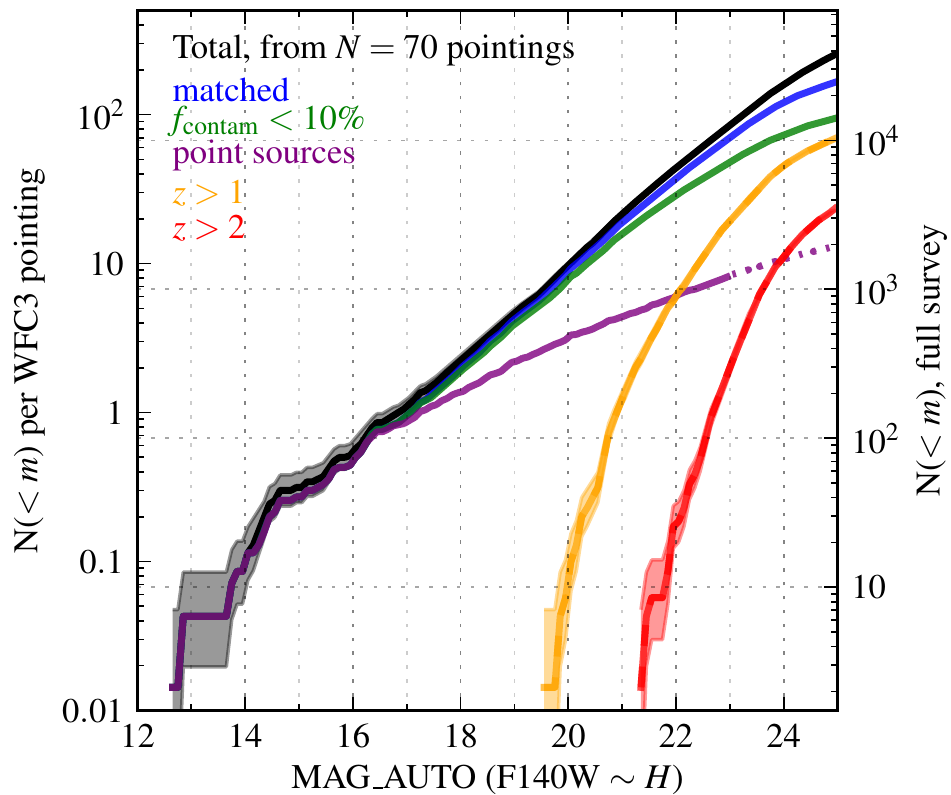}
\caption{Cumulative number counts, per WFC3 pointing.  The solid black line shows all objects detected in the direct images in 70 analyzed pointings from COSMOS, AEGIS, and GOODS-North/South fields.  The blue line shows the 3D-HST objects matched within 1 arcsec to an object in the ancillary photometric catalogs.  The green line shows the objects whose flux at 1.4\micront\ is contaminated at a level less than 10\% by flux from their neighbors.  The purple line shows objects identified as point sources based on their SExtractor sizes, which are clearly separated from extended sources down to $H_{140}=23$.  The orange and red lines show the number counts of galaxies with estimated redshifts $z>1$ and $z>2$, respectively, from the redshift fits described in Section \ref{s:redshift_fits}.  The shaded regions show Poisson-like 1$\sigma$ confidence intervals computed following \cite{gehrels:86}.
\label{f:number_counts}}	
\end{figure}

\begin{figure*}
\epsscale{0.95}
\plotone{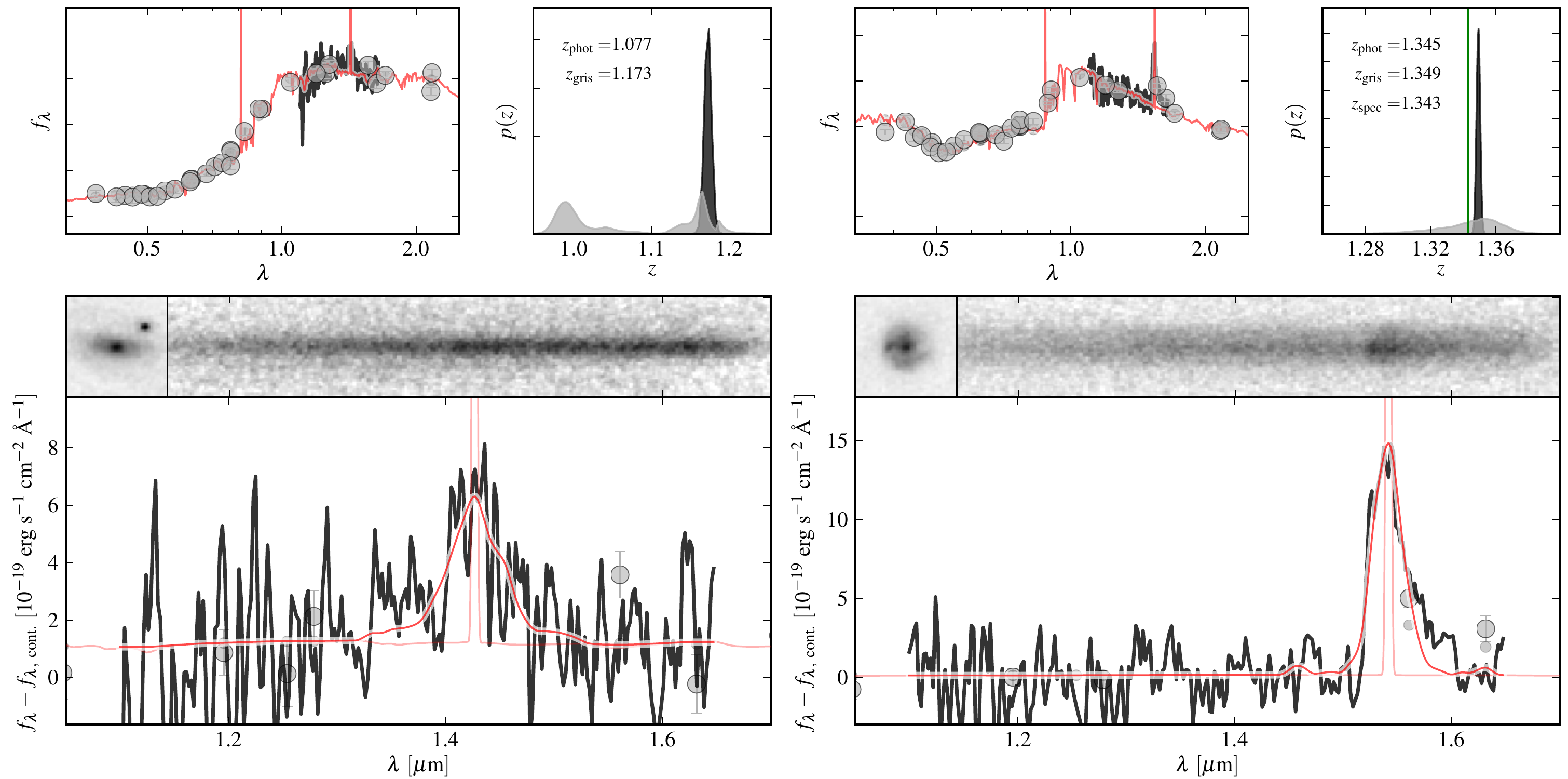}
\caption{Demonstration of the 3D-HST redshift fitting from two objects from the COSMOS field at $z>1$.  The top left panels show the full SED used in the fit, which includes the $U$---8\micront\ photometry and the F140W grism spectra.  The best-fit (unconvolved) template is shown in light blue.  The top right panels show the redshift probability distributions, where the light gray is for the photometry alone and the darker curve is for the fit including both the spectrum and the photometry.  The grism spectra greatly improve the redshift constraints compared to the photometry alone.  The best-fit redshift of the object at right differs from the independent spectroscopic measurement taken from \cite{lilly:07} (green line) by only $0.0025\times(1+z)$.    The bottom panels demonstrate how the spectral resolution of the slitless spectra is determined by the object profile.  The direct images are shown in the small inset panels and the 2D grism spectra are adjacent to the right.  The bottom panels show continuum-subtracted spectra along with the convolved and unconvolved line fits shown in red.  For the object on the right in particular, the 2D spatial distribution of the line emission closely follows that of the direct image, which is the sum of the line + continuum across the filter bandpass.  The convolved template fits are very good representations of the observed H$\alpha$+\ion{N}{2} line profiles.  The integrated line fluxes 2.8 and 4.8$\times10^{-16}\ \fluxcgs$ for the objects at left and right, respectively, and their equivalent widths are $43\pm4$ and $91\pm3$\ \AA.  Note that while the line is clearly detected by the fit for the object at left, it would be difficult to pick out the line emission ``by eye'' from the 2D spectrum, particularly in a blind search. 
\label{f:example_spectra}}	
\end{figure*}

Figure \ref{f:number_counts} shows the number of objects brighter than a given magnitude found in each 2$\times$2 arcmin WFC3 pointing of the 3D-HST survey.  The number counts shown are taken from 70 pointings taken as of 2011 August, and the number of a particular type of object expected in the full 149 pointings of the 3D-HST survey (including the GOODS-N pointings) is shown on the right axis.  The large majority of 3D-HST objects are matched within $1\farcs0$ to objects in the ancillary photometric catalogs ($>$90\% at $H<23$), though the fraction of matched objects begins to decrease significantly at fainter magnitudes where the shallower ancillary catalogs are incomplete.  The fraction of objects whose spectra are significantly contaminated by flux from their neighbors increases with magnitude because the object surface density increases and faint galaxies are more likely to be close to other faint galaxies and also because even the extended envelopes of nearby brighter galaxies can have fluxes similar to faint galaxies.  However, the fraction of largely uncontaminated objects is still $\sim$50\% even at the faint limits of the survey.  Objects with radii less than $\sim 0\farcs18$ (SExtractor 50\% \flux_radius) are clearly separated from resolved objects down to $H_{140}\sim23$, and these point sources (i.e., stars) represent the majority of the brightest objects.  The cumulative number counts of galaxies with estimated redshifts $z>1$ and $z>2$ (Section \ref{s:redshift_fits}) are also indicated in Figure \ref{f:number_counts}.  Based on the available pointings, some 7000 galaxies with $z>1$ and $H_{140} < 23.8$ are expected within the full survey.  For most magnitudes fainter than this practical continuum limit, the redshift estimates will converge to the limit of pure photometric redshifts limited by the depth of the ancillary photometry.  However, 3D-HST will provide precise redshifts and line fluxes for a smaller sample of galaxies with extreme equivalent widths, such as the star-bursting dwarf galaxies at $z>1.5$ discovered recently by \cite{vanderwel:11} (see also Figure \ref{f:eqw_mag}).

%
%
\section{Analysis}

The primary goal of the 3D-HST survey is to measure precise redshifts of a relatively unbiased sample of the galaxy population at $1 < z < 3$ to $H\approx23.8$.  In many cases, redshifts can be easily measured from emission lines, though only a single emission line will be observed by either the WFC3 or ACS grisms for most redshifts other than within a few narrow redshift intervals (see Figure \ref{f:throughput}).  The combination of the optical and IR grism spectra will provide multiple emission lines for many galaxies, though the sensitivity of the parallel ACS/G800L grism exposures is considerably less than that of WFC3/G141.  Redshifts measured from narrow absorption and broad continuum features \citep[e.g.,][]{vandokkum:10b} require that the systematic effects on the continuum shape be minimized, which is a particular challenge with the slitless grism spectra (i.e., removing the background and contamination of overlapping spectra).  For all of the reasons described above, the addition of multi-wavelength photometry is critical for the redshift and SED-fitting analysis of the grism spectra of the general galaxy population.

\subsection{Ancillary photometric catalogs}

The 3D-HST/CANDELS fields are among the best-studied extragalactic survey fields in the sky.  All of the fields have extensive multiwavelength observations from X-ray to radio wavelengths, the details of which can be found in \cite{grogin:11}.  For the redshift and SED fitting analysis of the grism spectra, we require deep optical and near-IR photometric catalogs selected at near-IR wavelengths, corresponding to the rest-frame optical for the redshifts of interest.  Currently we simply match the F140W-selected objects within the 3D-HST fields to the deepest K-selected catalogs available with broad multiwavelength coverage: AEGIS, COSMOS, \cite{whitaker:11};  GOODS-S, \cite{wuyts:fireworks}; GOODS-N, \cite{kajisawa:11}; and UDS, \cite{williams:09}.  These catalogs have between 8 (UDS) and 35 (COSMOS) photometric bands from $U$ through the \textit{Spitzer}-IRAC channels at 3--8\micront.  The image quality of the ground-based catalog detection bands ($0\farcs7$--$1\farcs0$) is not well matched to that of the 3D-HST imaging ($0\farcs13$).  As a result, multiple 3D-HST objects can be matched to single objects in the photometric catalogs (cf. $\sim$15\% at $H$=23), which, among other effects, can produce artificial 3D-HST pairs in the current analysis.  Furthermore, even the relatively short F140W exposures are somewhat deeper than the ground-based $K$-band catalog detection images (e.g., $K<22.8$ for the \citealp{whitaker:11} COSMOS catalog, see Figure \ref{f:number_counts}).  For these reasons, we are developing PSF-matched and deblended photometric catalogs, following \cite{labbe:06}, selected from the high-resolution, deep 3D-HST F140W and CANDELS images that will provide a one-to-one correspondence between objects with extracted spectra and their measured ancillary photometry (R. Skelton, in prep).

\subsection{Redshift and emission line fitting}\label{s:redshift_fits}

\begin{figure}
\epsscale{1.1}
\plotone{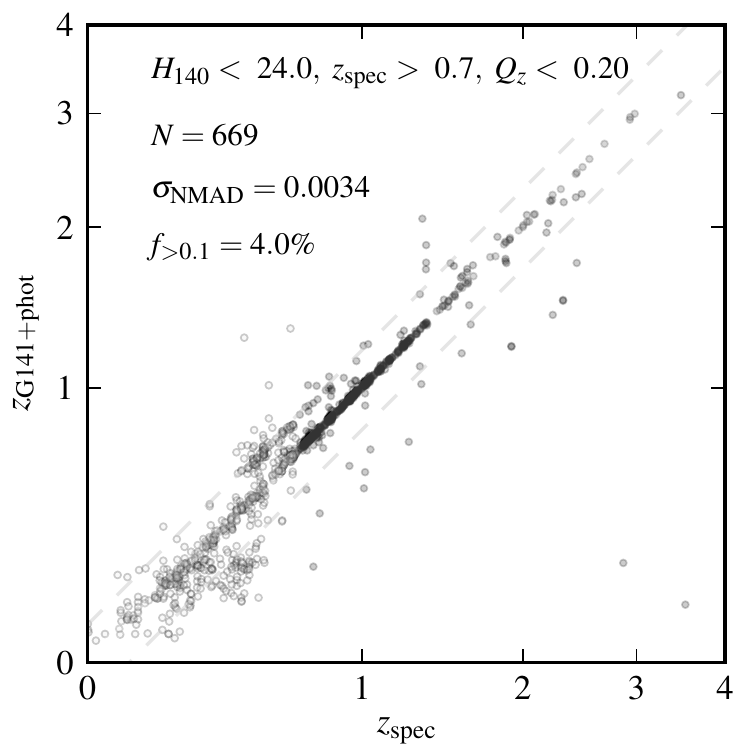}
\caption{Comparison of the 3D-HST redshift measurements to spectroscopic redshift measurements taken from the literature.  The 3D-HST sample is selected as indicated, with $H_{140} < 24$ and also including a cut on the \eazy\ redshift quality parameter $Q_z<0.2$, above which the quality of the template fits clearly decreases (among other parameters of the fit, $\chi^2$ is incorporated in $Q_z$, see \citealp{brammer:08}).  The redshift precision is $\sigma_\mathrm{NMAD}=0.0034(1+z)$ (see \citealp{brammer:08} for the definition of the normalized median absolute deviation, or NMAD) with 4\% catastrophic outliers at redshifts, $z>0.7$, where strong emission lines and continuum features fall within the coverage of the G141 grism.  The sources for the spectroscopic redshifts are as follows: AEGIS, \cite{davis:deep2, steidel:03}; COSMOS, \cite{lilly:07}, \cite{brusa:10}; GOODS-South, compilation from \cite{wuyts:fireworks}, \cite{balestra:10}; GOODS-North, compilation from \cite{barger:08}, \cite{steidel:03}, \cite{cooper:11}. 
\label{f:zphot_zspec}}	
\end{figure}

\begin{figure}
\epsscale{1.2}
\plotone{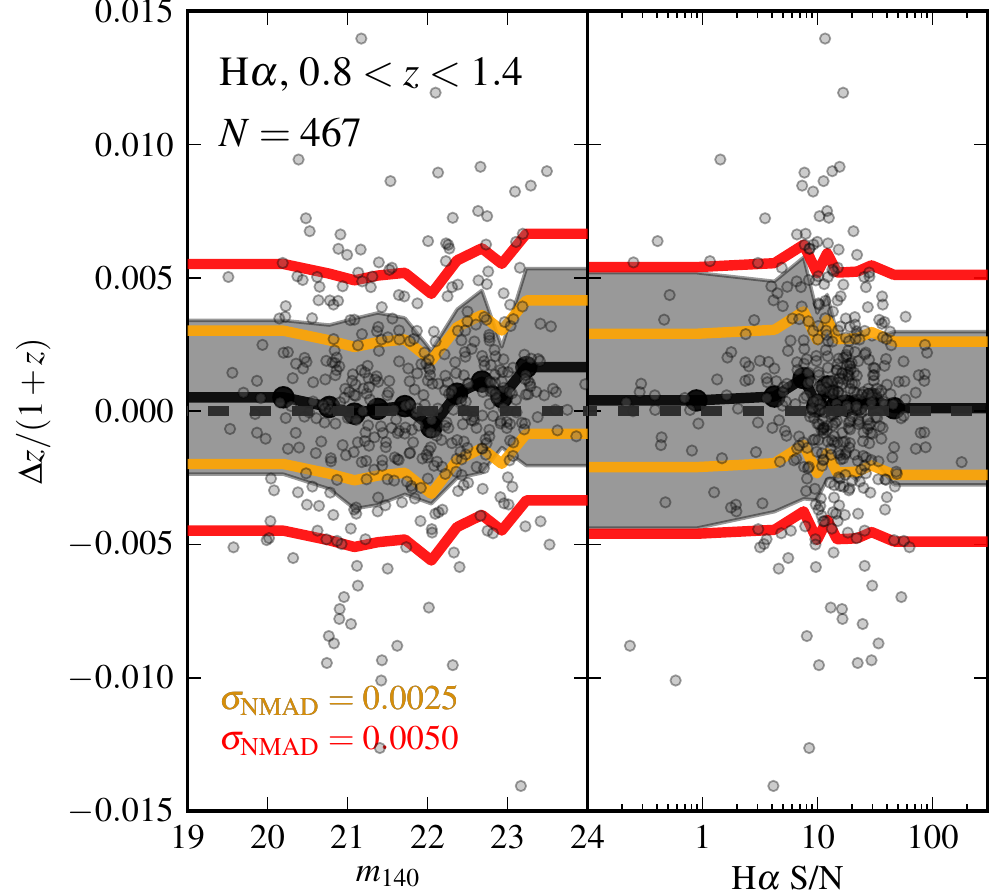}
\caption{Deviations of the grism redshifts from the previously-measured spectroscopic value as a function of magnitude (left panel) and signal-to-noise of the H$\alpha$ emission line (right panel).  The thick black lines shows the average deviation in bins with equal numbers of points ($N=46$) and the shaded gray region shows the observed NMAD scatter within these bins.  For comparison the solid orange and red lines indicate values of constant scatter of $\sigma_\mathrm{NMAD}=0.0025$ and 0.005, respectively, with respect to the median.  While there is little trend in the grism redshift offset or scatter as a function of magnitude, the scatter decreases significantly with increasing emission-line strength.
\label{f:zphot_zspec_lines}}	
\end{figure}

We estimate the redshifts of 3D-HST galaxies using a modified version of the \eazy\ code \citep{brammer:08}.  Two examples of the fits are shown in Figure \ref{f:example_spectra}.  The inputs to the fit are the 1D grism spectra, calibrated and extracted with \aXe, combined with broad-band and medium-band photometry from the catalogs described above when there is an object in the catalog matched within $1\farcs0$ of a 3D-HST object.  We subtract the quantitative contamination model directly from the spectra (Figure \ref{f:grism_model}).  Finally, we compute a normalization and a linear tilt to scale the G141 spectra to the $J$ and $H$ broad-band and medium-band photometry, in order to minimize residuals affecting the continuum shape that remain after the contamination correction and background subtraction.  At a minimum, a zeroth order normalization is necessary to match the flux-calibrated spectrum to the total photometric fluxes that have been measured within an effective infinite aperture \citep[see, e.g.,][]{whitaker:11}.

In the standard operation of the \eazy\ photometric redshift code, the model fluxes of the high-resolution spectral templates at a given redshift are computed by convolving the templates with the filter transmission curve of each photometric band.  We define effective ``filter curves'' for each grism spectral bin (of each object individually) by convolving a boxcar with the width of the spectral bin with the object profile.  This profile is taken from the direct image and averaged along the spectral dimension.  This accounts for both the possible tilt of the effective slit defined roughly along the object major axis, which is not necessarily perpendicular to the dispersion axis of the slitless spectra (Section \ref{s:spectra_extraction}), as well as higher-order features in the object profile (e.g., spiral arms).  In principle the template fitting could be done on the full 2D spectra directly, though we opt for the relative computational simplicity of using the extracted 1D spectra for the present analysis.  An implicit assumption of this method is that the 2D profile is ``gray'', i.e., that the profile is the same in the line as in the continuum across the spectral range covered by the grism.  Investigating  the validity of this assumption is in fact one of the primary goals of the 3D-HST survey \citep[see, e.g.,][]{nelson:12}, made possible by the superb spatial resolution provided by the \HST.

\begin{figure*}
\epsscale{1.1}
\plotone{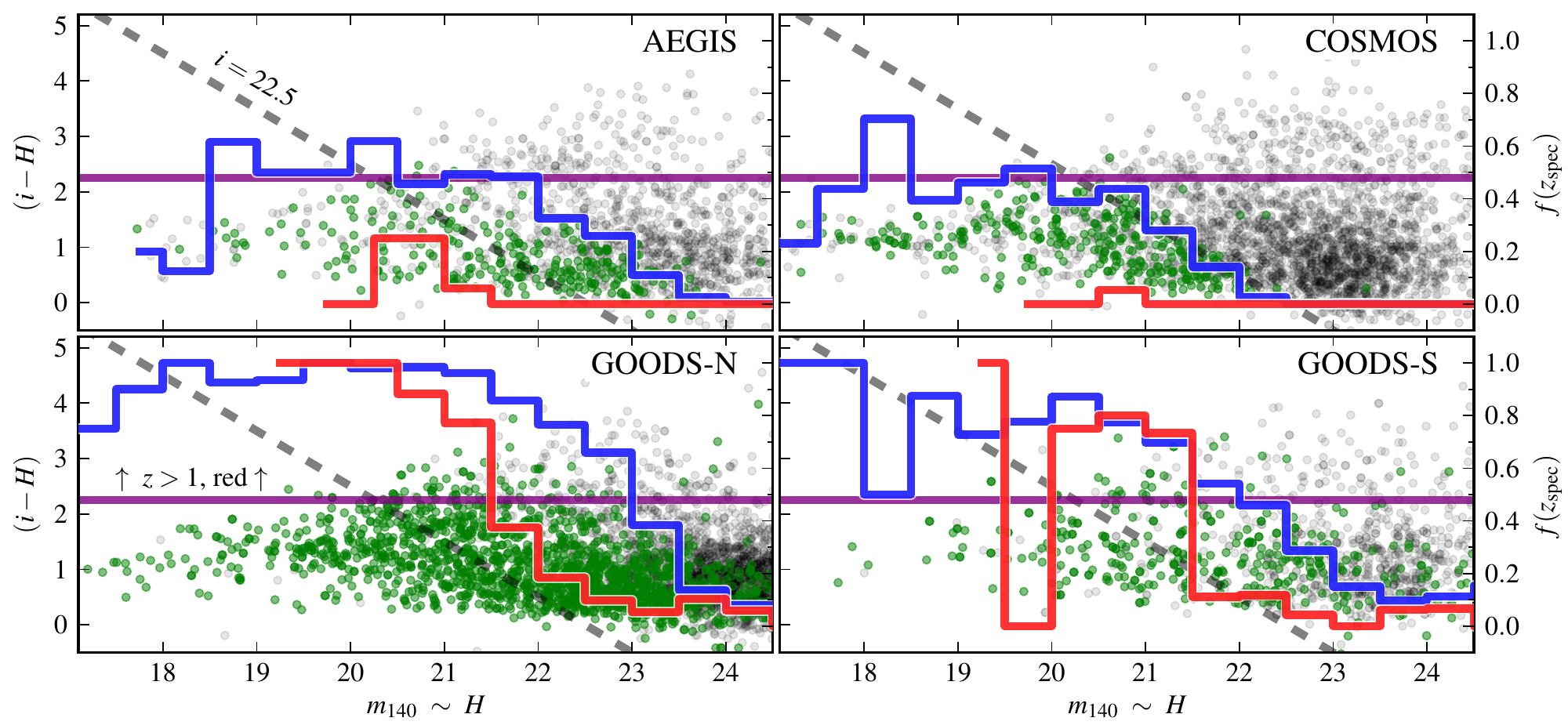}
\caption{Coverage of ground-based spectroscopic redshift measurements as a function of observed $(i-H)$ color and $i$ magnitude.  All galaxies in the available 3D-HST pointings are shown with gray symbols, and those marked in green have ground-based spectroscopic measurements taken from the references summarized in Figure \ref{f:zphot_zspec}.  The dashed lines in all panels indicate a constant $i=22.5$.  Analogous to the ``distant red galaxy'' selection of \cite{franx:03}, galaxies with $(i-H) > 2.2$ (purple lines) are nearly all at $z>1$.  The solid blue and red histograms indicate the fraction of galaxies with ground-based spectroscopic redshifts bluer and redder than this limit, respectively, with the scale shown on the axes at right.  \label{f:zspec_fraction}}	
\end{figure*}

The default set of the \eazy\ spectral templates contain emission lines added by determining an effective SFR for each template and then adding lines with empirically-calibrated line ratios (\citealp{brammer:11}, see also \citealp{ilbert:10}).  In order to fit the 3D-HST spectra + photometry we remove the emission lines from the galaxy templates and provide separate templates for emission lines of H$\alpha$, H$\beta$, [\ion{O}{3}]$\lambda$4959+5007, and [\ion{O}{2}]$\lambda$3727 individually.  The non-negative normalizations of the galaxy templates and emission lines are computed simultaneously by the code, providing an implicit measurement of the line strengths and equivalent widths along with the redshift fit.  Also by construction, the equivalent widths of the hydrogen Balmer lines are corrected for underlying absorption.  

For all but the most extreme velocity-broadened lines, the line shape is determined by the object profile.  For example, the spectral resolution of an object with (spatial) $\mathrm{FWHM}=0\farcs5$ is approximately 180\AA, or 1500 \kms for H$\alpha$ at $z=1.3$.  Due to the low spectral resolution, H$\alpha$ and \ion{N}{2}$\lambda$6550+6584 are not resolved and the H$\alpha$ line measurements represent the sum of these line species.  The G141 spectral resolution tends to be just sufficient to produce an asymmetrical profile for the [\ion{O}{3}]$\lambda$4959+5007 doublet, which can help in differentiating it from H$\alpha$ assuming that the profile of the line-emitting region is roughly symmetric (see also \citealp{atek:10, trump:11} for examples of H$\alpha$ and [\ion{O}{3}] line profiles).

A comparison of redshifts determined from the grism+photometry fits to ground-based spectroscopic redshift measurements taken from the literature is shown in Figure \ref{f:zphot_zspec}.  The precision of the grism redshifts is excellent ($0.0035\times(1+z)$ at $z>0.7$) and is an order of magnitude better than is typically possible with high-quality broad-band photometry alone (cf. $0.034\times(1+z)$, \citealp{brammer:08}).  Furthermore, the redshift precision is nearly constant over the full redshift range of interest $1 < z < 3$, whereas the precision of typical photometric redshifts is generally lower at these redshifts as relatively fewer observed photometric bands sample rest-frame optical wavelengths and the Balmer break \citep{brammer:11}.  

The precision of the redshift measurements depends somewhat on the strength and availability of emission lines in the grism spectra.  This is demonstrated in Figure \ref{f:zphot_zspec_lines}, which shows the deviations of the grism redshifts from the ground-based spectroscopic measurements as a function of F140W magnitude and the S/N of the H$\alpha$ emission line for objects at $0.8 < z < 1.4$ where the line lies within the coverage of the G141 grism.  While the redshift precision is nearly constant as a function of F140W magnitude, the redshift precision increases from only $\sigma\sim0.0025$ for galaxies with strong emission lines to $\sigma\sim0.005$ for galaxies without detected emission lines (S/N$\lesssim5$).  The sample shown in Figure \ref{f:zphot_zspec_lines} is dominated by objects from the extensive catalog of \cite{barger:08} in the GOODS-N field (see also Figure \ref{f:zspec_fraction}), which has fewer ancillary photometric bands than the other 3D-HST fields.  Therefore, it is notable that the precision of the continuum-only grism redshifts is still significantly better than has been achieved for surveys using many intermediate-width photometric filters \citep[e.g.,][]{wolf:04, ilbert:09, whitaker:11}.  

Of the galaxies shown in the comparison of Figure \ref{f:zphot_zspec}, some 4\% have redshift measurements that differ by more than $0.1\times(1+z)$ from the spectroscopic value.  The causes of these outliers include misidentification of single emission lines, poor background and/or contamination subtraction of the G141 spectra and apparent errors of the spectroscopic redshifts themselves, where a clear emission line (or lines) is observed in the grism spectrum and the overall SED (spectrum+photometry) is well fit by the \eazy\ templates.  In the current analysis, some significant systematic redshift errors remain at $z<0.7$, where the WFC3/G141 grism does not sample strong spectral features.  As we do not currently impose any constraints on the emission line ratios, these errors are typically caused by the fitting algorithm incorrectly adopting emission lines with extreme equivalent widths to fit optical continuum features sampled by individual, non-overlapping broad-band filters.  These effects will be removed by future improvements to the fitting code and, in particular, by the incorporation of the ACS/G800L optical grism spectra.

For the $z>0.7$ selection shown in Figure \ref{f:zphot_zspec},  30\% of the full 3D-HST sample have a measured ground-based spectroscopic redshift.  However, this ratio varies greatly as a function of object magnitude and color among the survey fields depending on the selection criteria of the spectroscopic surveys that cover them.  Figure \ref{f:zspec_fraction} shows the distribution of galaxies with measured spectroscopic redshifts as a function of $H$ ($m_{140}$) magnitude and observed $(i-H)$ color.  Spectroscopic surveys selected in the optical are most complete for bright, blue galaxies, typically at $z\lesssim1$.  For galaxies with $H<21$ and $(i-H) < 2.2$, 50\% in the COSMOS and AEGIS fields have a measured spectroscopic redshift and this fraction reaches 80\% and nearly 100\% for the deeper surveys of the GOODS-South and North fields.  At fainter magnitudes, optically-selected surveys only observe the bluest galaxies and do not sample the significant population of redder galaxies, typically at $z>1$.  Measuring spectroscopic redshifts for this population generally requires near-infrared spectra, which are challenging and time-consuming to obtain from the ground for large samples \citep[e.g.,][]{kriek:08a}.  3D-HST provides near-infrared spectra of all of the galaxies in Figure \ref{f:zphot_zspec}, which yield both crucial spectroscopic coverage of galaxies missing from typical redshift surveys and rest-frame optical spectra of galaxies at $z\gtrsim1$ that may have been observed previously only at rest-frame UV wavelengths \citep[see, e.g.,][]{vandokkum:11}.

\subsection{Example G141 grism spectra}

\begin{figure*}
\epsscale{1.18}
\plotone{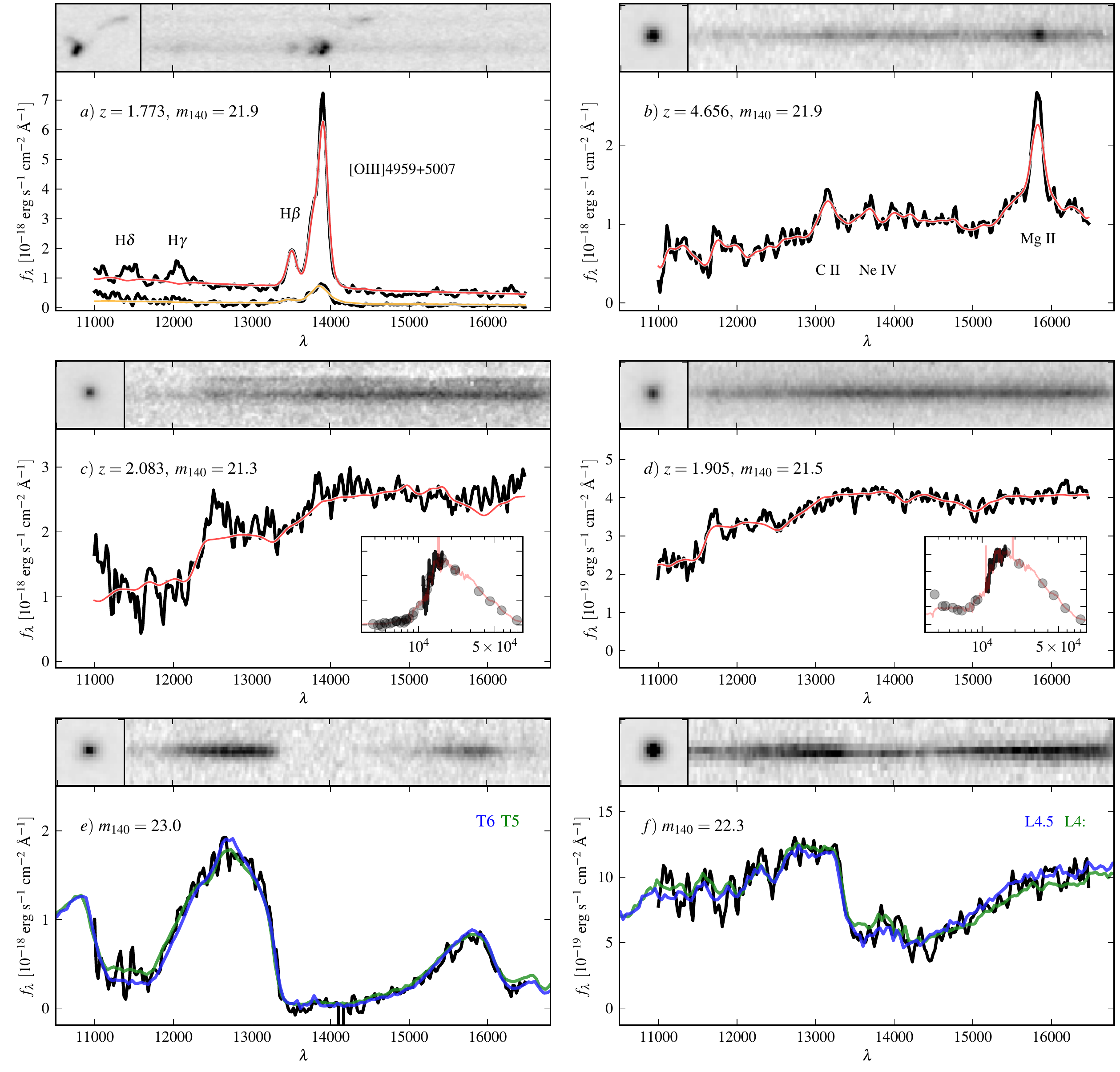}
\caption{Some of the diversity of objects within the 3D-HST survey. The template fits from the modified \eazy\ fits to the spectra + photometry are shown in the red and orange lines in panels $a$--$d$.  Panel \textbf{a)} shows an object in the GOODS-North field with multiple line-emitting components.  Two separate spectra are shown extracted for the bright compact component (which itself has two close sub-components) and the fainter, more diffuse tail extending to the upper right of the image thumbnail.  Panel \textbf{b)} is a quasar in the COSMOS field at $z=4.656$ with strong emission lines of \ion{Mg}{2} and \ion{C}{2}.  Panels \textbf{c)} and \textbf{d)} show extremely massive galaxies ($10^{11.5}$ and $10^{11.2}\ M_\odot$) at $z\sim2$ with strong continuum breaks and no visible emission lines.  The inset panels show the full 0.3--8\micront\ SEDs (photometry + spectra)  and the template fit.  The bottom panels \textbf{e)} and \textbf{f)} show the spectra of T- and L-type brown dwarf stars, found in the AEGIS and GOODS-N fields, respectively.  The two best-fitting spectral templates from \cite{burgasser:10} are plotted on top of the spectra, with the spectral types indicated.  We emphasize that while the selection of objects shown have spectra with particularly high signal-to-noise, none of these objects are ``serendipitous'' detections:  that 3D-HST provides high-quality near-IR spectra of a wide variety of classes of objects is the very essence of the survey.
\label{f:object_examples}}	
\end{figure*}

As a blind survey of every object that falls within a given WFC3 (and/or ACS) grism pointing, 3D-HST produces high quality spectra for a broad diversity of celestial objects.  Some examples of the NIR grism spectra, chosen not as a representative sample but rather to demonstrate some of this diversity are shown in Figure \ref{f:object_examples}:  

\begin{itemize}
	
\item Figures \ref{f:object_examples} $a)$ and $b)$ show examples of line-emitting galaxies.  The object in panel $a)$ has multiple components that are all strong [\ion{O}{3}]$\lambda$4959+5007 emitters with large [\ion{O}{3}]/H$\beta$ line ratios.  The spatial extent of the [\ion{O}{3}]$\lambda$4959+5007 is clearly visible in the 2D grism spectra and closely follows the spatial profile in the direct image.  A typical magnitude-limited spectroscopic survey would likely only be able to place a slit on one of the two objects and their physical association would be unknown.  There are many such line-emitting pairs and groups within the 3D-HST survey that will be used to study the merger history of the universe out to $z\sim2$.  

Panel $b)$ shows a quasar in the COSMOS field at $z=4.656$, identified from strong emission lines of \ion{Mg}{2} and \ion{C}{2} and a strong Lyman break in the ancillary optical photometry (not shown).  The template fit of the rest-frame UV continuum is also in remarkable agreement with the G141 spectrum, showing numerous low-level absorption features.  The galaxy is identified as an X-ray source by \cite{elvis:09}, but its distant redshift has not otherwise been identified; it has $z_\mathrm{phot}=0.48$ in the catalog from \cite{ilbert:09} and is too faint ($i=22.6$) to be included in the zCOSMOS-bright sample \citep{lilly:07}.  In this sense, among other things 3D-HST is a near-IR spectroscopic survey of X-ray (or radio, or 24\micront, etc.) selected sources within the survey fields. 

\item Figures \ref{f:object_examples} $c)$ and $d)$ show NIR spectra of extremely massive ($M>10^{11.2}\ M_\odot$) galaxies in the COSMOS and GOODS-S fields. The grism spectra covering the Balmer/4000~\AA\ break, even with their low resolution, can greatly improve the constraints on the star formation histories of massive galaxies such as these \citep{vandokkum:10b}.  The remarkable galaxy in panel $d)$ shows a strong 4000 \AA\ break characteristic of a significantly evolved stellar population. The object in panel $d)$ is taken from the CANDELS grism pointing within the UDF  \citep[see][]{trump:11}.  The spectrum is somewhat deeper ($t_\mathrm{exp}=15$ ks) than typical 3D-HST G141 spectra, but was fully reduced with the 3D-HST pipeline.  

\item The bottom two panels of Figure \ref{f:object_examples} show spectra of two more nearby objects, cool brown dwarf stars with spectral types T5/6 and L4.  The spectral types are estimated from template fits from the SpeX spectral library \citep{burgasser:10}, with the two best fits shown plotted along with the spectra.  The T5/6 dwarf of panel $e)$ is the faintest (i.e., most distant) T-dwarf by almost three magnitudes in $H$ (F140W)\footnote{Compared to the compilation at \scriptsize{\url{http://www.dwarfarchives.org}}.  \footnotesize Note also that the T dwarf has $H_\mathrm{tot}=22.9$ in the more standard $H$ filter from the CFHT WIRDS photometry of the AEGIS field \citep{bielby:11}.}.  The T5/6 dwarf has a photometric distance of 300--400 pc using the $M_\mathrm{H}$ relations tabulated by \cite{vrba:04}, with the broad range resulting from the crude estimate of its spectral type.  Recently, three similarly faint, distant, cool brown dwarfs have also been found in the WISP grism survey by \cite{masters:12}.

\end{itemize}

Admittedly, the objects shown in Figure \ref{f:object_examples} are somewhat brighter than typical galaxies found in the full 3D-HST survey.  We provide additional example spectra in Appendix \ref{s:fainter_examples} down to the effective sensitivity limits of the survey.  Continuum breaks can indeed be detected down to the continuum limit ($H_{140}\sim23$), though the redshift precision of such faint, line-free, continuum objects will likely be intermediate between that shown in Figure \ref{f:zphot_zspec_lines} and that of medium-band photometric surveys such as the NEWFIRM Medium Band Survey (see Figure \ref{f:zphot_zspec_lines}). Robust 5$\sigma$ emission line detections are found at considerably fainter magnitudes at line fluxes consistent with the limits described in Section \ref{s:grism_sensitivity}.  For more examples of 3D-HST grism spectra, see \cite{vandokkum:11} for the one-dimensional spectra of a complete, physically defined, sample of massive galaxies at $1 < z < 1.5$, and a sample of H$\alpha$ emission line morphologies is presented by \cite{nelson:12}.

%
%
\section{Science objectives of the survey}\label{s:goals}

The 3D-HST survey described above is ideally-suited for studying galaxy evolution in the key epoch $1 < z < 3.5$, providing an important complement to the large multi-cycle CANDELS imaging program \citep{grogin:11, koekemoer:11} by measuring the crucial third dimension, redshift, for some 10$^4$ galaxies (Figure \ref{f:number_counts}).  \cite{grogin:11} summarize much of the science that will be enabled by a large \HST near-IR imaging program.  In a discussion that is by no means exhaustive, we describe below some of the science questions that require both high resolution imaging and the unique spectra that currently only 3D-HST can provide.

\subsection{What causes galaxies to stop forming stars?}

In the low redshift universe many galaxies are observed to be quiescent, with current SFRs only $\sim$1\% of their past average \citep[e.g.,][]{pasquali:06}.  These quiescent galaxies tend to be massive early-type galaxies, forming the ``red sequence'' in the color-mass distribution of galaxies.  Recent work has shown that at $z\sim2$ many massive galaxies ($M \gtrsim 10^{11}\ M_\odot$) exhibit spectacularly high SFRs of hundreds of solar masses per year, whereas others were already quiescent, particularly those that are extremely compact for their mass \citep{kriek:06b, vandokkum:08, brammer:09}.  Active galactic nucleus (AGN) feedback is a possible mechanism to suppress gas cooling and star formation \citep[e.g.,][]{croton:06}, but direct evidence is scarce.  

Diagnostics of quiescence can be correlated with stellar mass, surface density (i.e., compactness), and the environment of galaxies on Mpc scales.  These diagnostics are most reliably identified spectroscopically, through the strength of the Balmer- or 4000 \AA\ break ($D_{4000}$; see Figure \ref{f:object_examples}$c$,$d$) and/or the absence of emission lines such as H$\alpha$.  For a  5$\sigma$ limiting emission line flux of $5\times10^{-17}\ \fluxcgs$ (Section \ref{s:grism_sensitivity}), 3D-HST will reach H$\alpha$ and [\ion{O}{2}]$\lambda$3727 SFR limits \citep{kennicutt:98} of $2.5$ and $25$ $M_\odot\,\peryr$ at $z=1$ and $z=2$, respectively. If the simultaneous presence of quiescent and star-bursting massive galaxies at $z\sim2$ is the result of their surface density or their environment, correlations should exist between the SFR and these parameters.  

\subsection{To what extent are galaxies shaped by their environment?}

The morphology--density relation \citep{dressler:80} states that ``early-type'' galaxies (mostly massive quiescent galaxies) are relatively abundant in dense environments such as groups and clusters.  However, at low redshift most massive galaxies are quiescent regardless of environment \citep{kauffmann:03, balogh:04}, so it is difficult to determine whether the environment provides a physical mechanism that alters the galaxies (e.g., through gas stripping), or whether dense environments simply are the place where quiescent galaxies tend to end up.  There remains some tension in the recent literature whether galaxy-mass-driven effects \citep{peng:10} or the properties of their host dark matter halos \citep{wake:12b} are the dominant factors shaping the star formation histories of galaxies.

To further disentangle the roles of mass and environment, epochs should be considered when ``massive'' did not yet directly imply ``quiescent'' \citep{vandokkum:11}.  The 3D-HST sample will be sufficiently large to determine the relation between SFR and environment in bins of fixed mass and redshift.  If no environmental dependence is observed at fixed mass, then the relations between galaxy properties and environment are simply a by-product of the underlying relations of both quantities with mass.  Even excellent photometric redshifts with errors $\delta z \approx 0.04(1+z)$ have a radial error on the comoving distance of $>150$ Mpc at $z=2$, which is larger than the distance of the Milky Way to the Coma cluster.  With redshift errors more than an order of magnitude smaller (Figure \ref{f:zphot_zspec}), 3D-HST will be able to provide a sensible definition of the environmental galaxy density on the scale of a few co-moving Mpc, as well as spectroscopic diagnostics of galaxy ``quiescence''. 

\subsection{How did disks and bulges grow?}

The epoch $1 < z < 3$ saw the wholesale transition from small, star-forming clumps evident in the deepest Hubble images to the ordered ``realm of galaxies'' seen today \citep[e.g.,][]{elmegreen:07, wuyts:12}.  Since $z\sim1$, most star formation has taken place in large spiral disks, but different modes may have been prevalent at earlier times, such as disks made of star-bursting clumps that coalesce to form compact spheroids directly \citep{dekel:09b}.

If bulges formed before disks, then objects with stellar masses 1--5$\times10^{10}\ M_\odot$ and star formation concentrated on 1 kpc scales should be seen at $z\sim3$.  These should evolve into young compact bulges surrounded by disk-like ($\sim$5 kpc) star formation at $z\sim2$, and then to old bulges and regular disks by $z\sim1$.  By contrast, if bulges formed mostly from subsequent merging of disks, extended star-forming disks should be pervasive at $z\sim3$.  The relative ages of the subcomponents of galaxies and the spatial extent of line-emitting star-forming regions can be measured from the spatially-resolved \HST grism spectra \citep{vandokkum:10b}.  Ground-based integral-field spectroscopy has demonstrated the power of spatially-resolved spectroscopy for galaxies at $z>1$ \citep[e.g.,][]{forster:09, law:09}, but it has been typically limited to more luminous, rare objects or relatively small samples.

\subsection{What is the role of mergers in galaxy formation?}

Although the merger-driven growth of massive galaxies is a common prediction of galaxy formation models, it has been difficult to test at higher redshift where mergers should be most common \citep[e.g.,][]{guo:08}.  The merger rate can be determined from physical pair statistics, but these have been difficult to measure due to insufficient spatial resolution and contamination by chance superpositions of unassociated galaxies \citep[e.g.,][]{williams:11, man:11}.  

With its spatial resolution of $0\farcs13$ and spectral resolution of $\delta v \approx 1000$ \kms (Figure \ref{f:zphot_zspec}), 3D-HST can spectroscopically identify true physical pairs and groups down to separations $\lesssim 5$ kpc, weeding out projected galaxy pairs (see also Figure \ref{f:object_examples}$a$).  Within the 3D-HST sample, the pair fraction can be measured as a function of mass and redshift, and the fractions can be turned into a merger rate using models \citep[e.g.,][]{kitzbichler:08, lotz:11, williams:11}.  The mass growth of galaxies due to mergers can be compared to the growth due to star formation, and the sizes, densities, and AGN content of the merger components can be determined.  The SFRs of the spectroscopically-identified merging pairs can be compared to predictions of hydrodynamical simulations, such as that mergers should play a large role in driving star formation activity and black hole accretion \citep[e.g.,][]{cox:06}.

%
%
\section{Summary}\label{s:summary}

In this paper, we present the 3D-HST survey, a 248-orbit Treasury program to obtain low resolution ($R\sim130$) slitless grism spectroscopy of $\sim$7000 galaxies at $z>1$ with the \HST grisms.  Combined with the grism coverage of the GOODS-North field, 3D-HST will survey 625 arcmin$^2$ of well-studied extragalactic survey fields (AEGIS, COSMOS, GOODS-N, GOODS-S, and UKIDSS-UDS), with two orbits of primary WFC3/G141 coverage across the survey and two to four orbits with the ACS/G800L grism in parallel.  Short ``direct'' images are taken in the WFC3/F140W and ACS/F814W filters to provide the wavelength reference for the grism spectra.  These images are also scientifically useful as they reach depths comparable to deep ground-based surveys ($H_{140} \lesssim 26.1$) with $\sim0\farcs13$ spatial resolution and sample wavelengths between the F125W and F160W images taken by the CANDELS survey.  

We have developed a custom data reduction pipeline built around the Multidrizzle and \aXe\ software tools, which can quickly and automatically reduce any typical direct + grism image sequence.  Of particular importance to the grism data reduction is the subtraction of the background, which comes from the combination of zodiacal light, Earth glow, and low-level thermal emission in the IR.  We demonstrate a method to subtract the variable structure in the background that offers significant improvement over the reduction with the default grism calibration files.  The primary products of the pipeline are calibrated 2D and 1D spectra extracted with \aXe.  We build a quantitative contamination model based on the 3D-HST direct images and multi-band imaging from the CANDELS survey to account for the fact that the spectra of nearby objects can overlap due to the lack of slits.  

The 3D-HST survey fields provide a wealth of ancillary multi-wavelength observations that are crucial for interpreting the grism spectra, which frequently only contain a single emission line, if any.  We adapted the \eazy\ code \citep{brammer:08} to measure redshifts from SEDs composed of both the grism spectra and matched photometry spanning 0.3--8\micront.  These spectrophotometric redshifts are more than an order of magnitude more precise than typical broad-band photometric redshift estimates, with $\sigma = 0.0034(1+z)$.  The redshift-fitting code also provides measurements of emission line fluxes and equivalent widths, which \cite{vandokkum:11} use to show that there is a broader diversity of H$\alpha$ line strengths (i.e., star formation activity) among massive galaxies at $1 < z < 1.5$ compared to those with similar masses locally.

3D-HST provides a spectrum for essentially every object in the field (to a magnitude limit, modulo contamination from overlapping spectra).  We demonstrate some of the diversity of objects found within the survey---from high-$z$ quasars to brown dwarf stars---which should not be considered ``contaminants'' or ``interlopers'' but are rather natural components of the survey.  The survey is optimally designed, however, for the study of galaxy formation over $1 < z < 3.5$.  Some of the science objectives that require the unique combination of high spatial resolution, deep near-IR (and supporting optical) spectra include disentangling the processes that regulate star formation in massive galaxies, evaluating the role of environment and mergers in shaping the galaxy population, and resolving the growth of disks and bulges, spatially and spectrally.

All of the raw WFC3 and ACS data for 3D-HST are immediately made public in the \HST archive, and we plan to release both low- and high-level data products within 18 months of the completion of the observations.  These data products will include calibrated 2D and extracted 1D spectra, ancillary photometric catalogs matched to the 3D-HST sample, and catalogs of physical parameters derived from the spectra and photometry such as redshifts, stellar masses and SFRs.  Much of the current reduction code is made available immediately at \url{http://code.google.com/p/threedhst/}, to which contributions and modifications from the community are encouraged.  News updates and data releases from the survey will be provided on the 3D-HST webpage at \url{http://3dhst.research.yale.edu/}.

%
%
\acknowledgements

We are grateful to the authors and maintainers of the \aXe\ software package: Martin K\"ummel, Harald Kuntschner, Jeremy Walsh, and Howard Bushouse.  This research has made extensive use of NASA's Astrophysics Data System Bibliographic Services and of open source scientific Python libraries, including PyFITS and PyRAF produced by the Space Telescope Science Institute, which is operated by AURA for NASA.  The brown dwarf spectral templates are taken from the SpeX Prism Spectral Libraries, maintained by Adam Burgasser at \url{http://pono.ucsd.edu/~adam/browndwarfs/spexprism}.  3D-HST is supported in part by grant HST-GO-12177 awarded by the Space Telescope Science Institute.  Additional funding for this research was provided by the Marie Curie Actions of the European Commission (FP7-COFUND) and the
European Research Council under the European Community's Seventh
Framework Programme (FP7/2007-2013) / ERC grant agreement 227749.


\bibliographystyle{apj}
\bibliography{ms}

%
%


\begin{appendix}

\section{Grism flat-field correction}\label{ap:flat_field}

\begin{figure}
\epsscale{1.}
\plotone{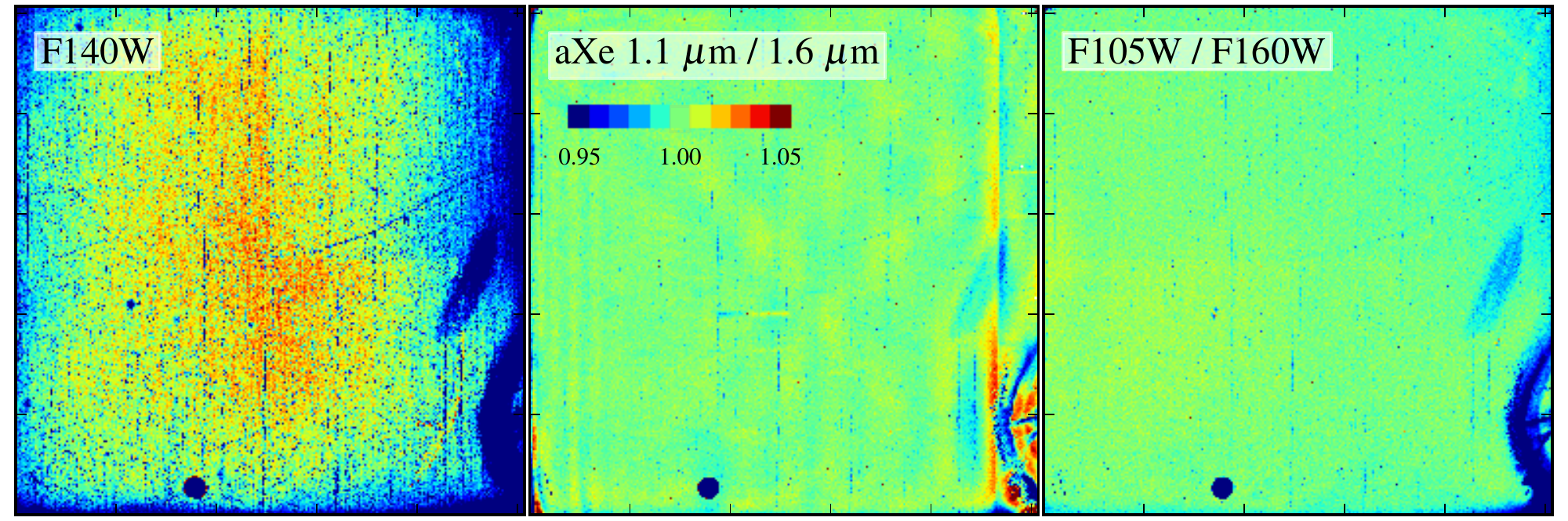}
\caption{Wavelength dependence of the WFC3/IR flat-field.  The left panel shows the F140W imaging flat-field, divided by the pixel area map to better show the pixel-to-pixel variations across the entire detector.  The middle panel shows the ratio of the wavelength-dependent G141 flat-field as used by aXe, calculated at 1.1\micront\ and 1.6\micront.  The right panel shows the ratio of the pipeline imaging flat fields for the F105W and F160W filters.  The color scale shown in the center panel is the same for the other panels. \label{f:wavelength_flats}}	
\end{figure}

Each pixel in a grism image ``sees'' the superposition of flux at different wavelengths coming from nearby positions on the sky where the grism disperses the flux to fall on top of that pixel.  The standard \aXe\ data reduction tries to account for this by applying a wavelength-dependent flat-field correction to pixels within the dispersed spectrum of a given object.  We adopt a more simplified treatment of the flat-field by simply dividing by the F140W imaging flat before the background subtraction and spectral extraction (Section \ref{s:background_subtraction}).  In doing so, we also separate multiplicative flat-field and additive background terms from the background subtraction.  However, this technique ignores any wavelength dependence of the flat-field, which is shown in Figure \ref{f:wavelength_flats}.  

The left panel of Figure \ref{f:wavelength_flats} shows the F140W flat itself, with the  large-scale variation resulting from the variable pixel areas\footnote{\scriptsize{\url{http://www.stsci.edu/hst/wfc3/pam/pixel_area_maps}}} removed for display purposes.  The center panel shows the estimated ratio of the flat-field at the blue to red edges of the G141 sensitivity, computed from the wavelength-dependent flat-field images used by \aXe.  The right panel of Figure \ref{f:wavelength_flats} shows the ratio of the pipeline imaging flats for the F105W and F160W filters.  
For both estimates, the wavelength dependence of the flat-field across the G141 sensitivity is generally less than $\pm$1\% outside of the ``wagon-wheel'' feature in the lower right corner of the detector.  The color dependence of the \aXe\ flat-field includes some high-frequency structure not seen in the ratio of the imaging flats, and also appears to underestimate the decreased sensitivity in the wagon wheel at blue wavelengths.  Our simplified treatment of dividing by the single F140W flat-field can therefore result in flat-field errors of $\sim$5\% in this part of the detector, however, dividing by this ``average'' flat is generally sufficient to flatten the background and reduce systematic effects caused by the background subtraction (see Figure \ref{f:subtract_background}).

\clearpage

\section{Background variations}\label{ap:background_variation}

\begin{figure}
\epsscale{0.5}
\plotone{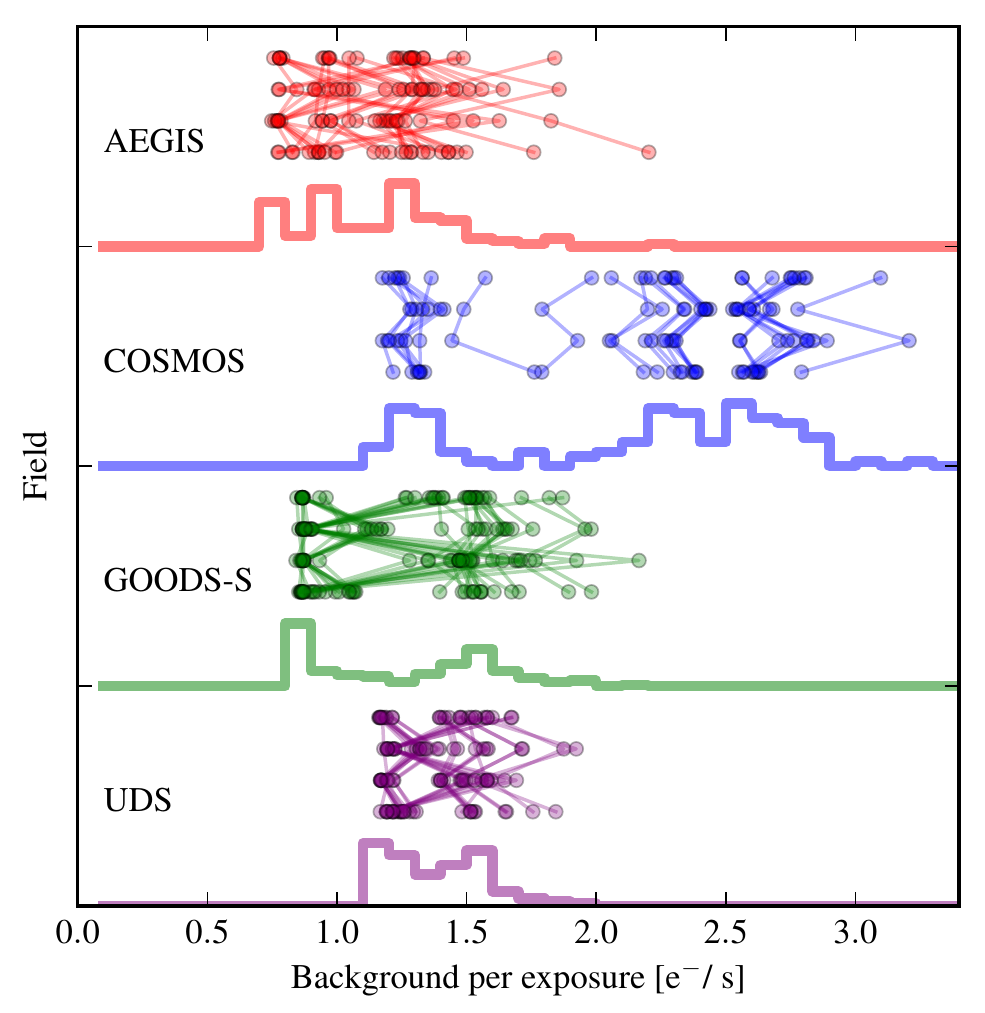}
\caption{Background levels for all of the individual exposures in the four primary 3D-HST survey fields.  The four exposures within a given visit are shown with circles connected by lines, showing that the background can vary significantly within a single visit (e.g., in GOODS-S).  Histograms indicate the overall distribution of background levels in each field. \label{f:background_variation}}	
\end{figure}

The background level in the 3D-HST grism exposures varies significantly from 0.5 electrons/s to 3 electrons/s depending on the field and the relative orientation of the instrument with respect to the zodial light and Earth glow sources of background emission.  The orientation of the bright limb can also vary within a single orbit, resulting in different overall background levels and also, occasionally, different 2D structure within the background (Section \ref{s:background_subtraction}).  Summarized in Table \ref{t:fields}, we present here the full distribution of background levels for all of exposures in the four primary 3D-HST survey fields in Figure \ref{f:background_variation}.  The COSMOS field shows the largest variation, with a majority of the pointings having background levels in excess of 2 electrons / s.  There are a number of GOODS-S pointings in which the background levels vary by roughly a factor of two (0.8--1.5 electrons / s) within a single visit.

\section{Example spectra of typical objects in 3D-HST}\label{s:fainter_examples}

\begin{figure*}
\epsscale{1.2}
\plotone{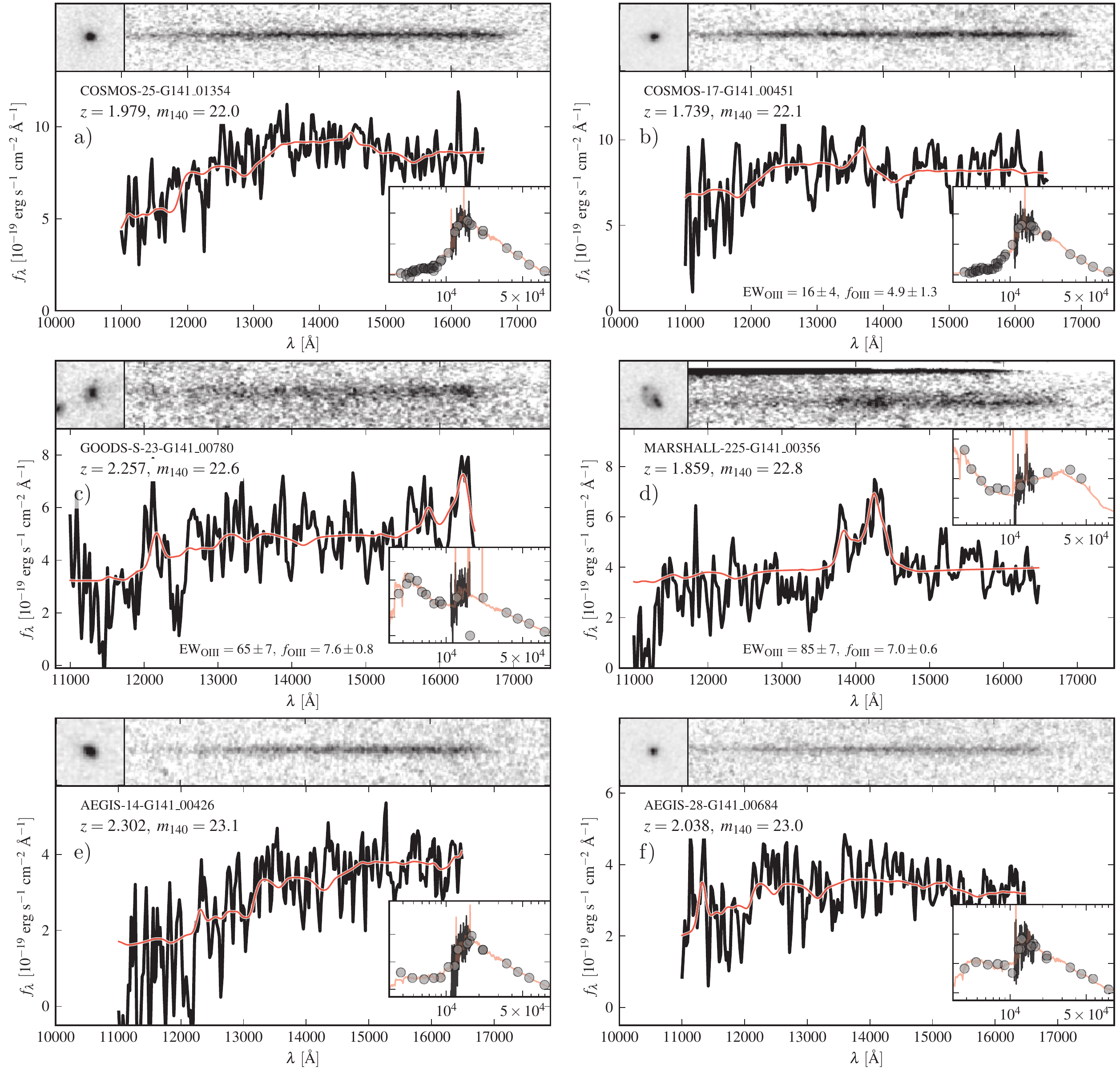}
\caption{Example spectra of more typical (i.e., fainter) galaxies at $z>1.7$ than those shown in Figure \ref{f:example_spectra}.  The components of each individual object plot are as described in Figure \ref{f:example_spectra}. The galaxies shown here have well-detected continuum (Balmer/4000\AA) breaks and weak, if any, emission lines.  The pointing where each object is found is indicated in the inset labels (``MARSHALL'' is a CANDELS supernova pointing in the UDS).  Where emission lines of \ion{O}{3}$\lambda$4959+5007 are clearly detected, the combined equivalent width and integrated flux of the doublet is indicated, in units of \AA\ and $10^{-17}~\fluxcgs$, respectively.\label{f:fainter_cont_examples}}	
\end{figure*}

\begin{figure*}
\epsscale{1.2}
\plotone{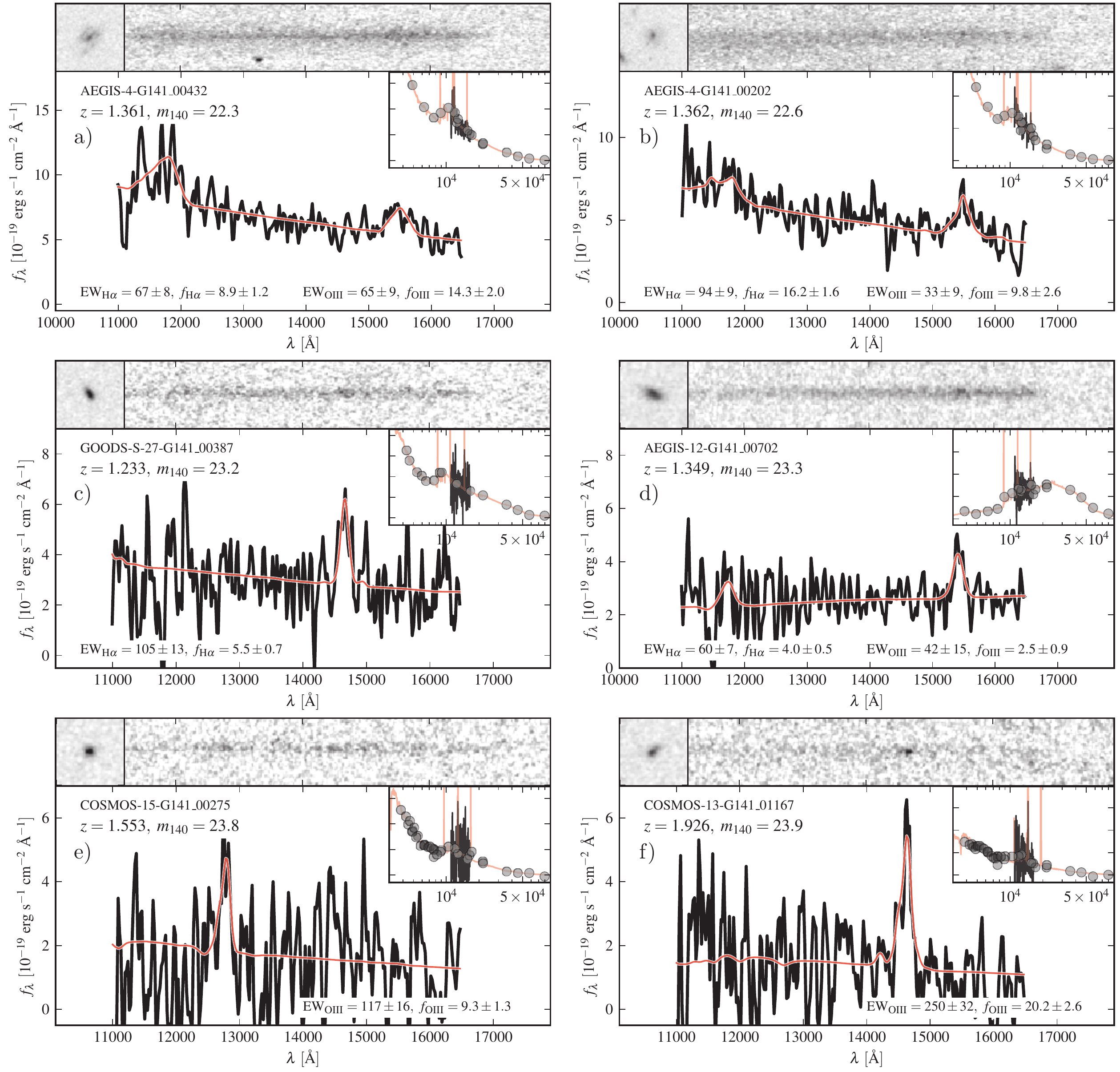}
\caption{Example spectra of line-emitting galaxies where the line signal to noise, $3 < {\rm S/N} < 8$.  The components of each individual object plot are as described in Figure \ref{f:example_spectra}.  The equivalent widths and integrated line fluxes of H$\alpha$ and \ion{O}{3}$\lambda$4959+5007 emission lines are indicated, in units of \AA\ and $10^{-17}~\fluxcgs$, respectively. \label{f:fainter_line_examples}}	
\end{figure*}

The objects with spectra shown in Figure \ref{f:example_spectra} are shown to demonstrate the remarkable quality of spectra that can be achieved with the WFC3/G141 grism for moderately bright targets (though even those magnitudes are currently difficult to reach from the ground).  However, the objects in Figure \ref{f:example_spectra} are brighter than the \textit{typical} object that will be found in the full 3D-HST survey (e.g., Figure \ref{f:number_counts}).  Here we provide a gallery of spectra at continuum (Figure \ref{f:fainter_cont_examples}) and emission-line (Figure \ref{f:fainter_line_examples}) brightnesses extending to the faint limits probed by the survey (Section \ref{s:grism_sensitivity}).  It bears noting that there appear to be significant narrow absorption features in some of the continuum spectra shown in Figure \ref{f:fainter_cont_examples}.  While the grism is capable of resolving absorption lines \citep[e.g.,][]{vandokkum:10b}, the narrow features in the faint spectra shown in Figure \ref{f:fainter_cont_examples} are more likely the result of correlated pixel noise that arise from the drizzling process.  It is also possible that such correlated noise could cause spurious \textit{emission} line detections, but the fact that the line shape predicted by the object morphology is in good agreement with the observed line shape in Figure \ref{f:fainter_line_examples} demonstrates that, at least for the objects shown, even faint lines can be robustly identified.

\end{appendix}

\end{document}